\documentclass[10pt]{article}
\usepackage{amsfonts,amsmath,amssymb}

\numberwithin{equation}{section}

\newcommand{\bR}{{\mathbb R}}

\newcommand{\bC}{{\mathbb C}}

\newcommand{\bN}{{\mathbb N}}

\newcommand{\kB}{{\mathcal B}}
\newcommand{\kC}{{\mathcal C}}
\newcommand{\kD}{{\mathcal D}}

\newcommand{\kH}{{\mathcal H}}

\newcommand{\kM}{{\mathcal M}}
\newcommand{\kN}{{\mathcal N}}

\newcommand{\kS}{{\mathcal S}}

\newcommand{\gotH}{{\mathfrak H}}

\newcommand{\gotS}{{\mathfrak S}}

\newcommand{\ga}{{\alpha}}

\newcommand{\gd}{{\delta}}
\newcommand{\gD}{{\Delta}}
\newcommand{\gga}{{\gamma}}
\newcommand{\gG}{{\Gamma}}

\newcommand{\gL}{{\Lambda}}
\newcommand{\Gl}{{\Lambda}}
\newcommand{\gl}{{\lambda}}

\newcommand{\gO}{{\Omega}}
\newcommand{\gP}{{\Pi}}

\newcommand{\gs}{{\sigma}}

\newcommand{\gth}{{\theta}}
\newcommand{\gT}{{\Theta}}

\newcommand{\slim}{\,\mbox{\rm s-}\hspace{-2pt} \lim}

\newcommand{\real}{{\Re{\mathrm e\,}}}
\newcommand{\imag}{{\Im{\mathrm m\,}}}
\newcommand{\dom}{{\mathrm{dom\,}}}
\newcommand{\ran}{{\mathrm{ran\,}}}

\newcommand{\tr}{{\mathrm{tr}}}

\newcommand{\clo}{{\mathrm{clo}}}

\newcommand{\spa}{{\mathrm{span}}}
\newcommand{\clospa}{{\mathrm{clospan}}}

\newcommand{\wh}[1]{{\widehat{#1}}}

\newtheorem{thm}{Theorem}[section]
\newtheorem{prop}[thm]{Proposition}
\newtheorem{lem}[thm]{Lemma}
\newtheorem{cor}[thm]{Corollary}

\newtheorem{defn}[thm]{Definition}
\newtheorem{rem}[thm]{Remark}

\newcommand{\ba}{\begin{array}}
\newcommand{\ea}{\end{array}}
\newcommand{\bea}{\begin{eqnarray}}
\newcommand{\eea}{\end{eqnarray}}
\newcommand{\bead}{\begin{eqnarray*}}
\newcommand{\eead}{\end{eqnarray*}}
\newcommand{\be}{\begin{equation}}
\newcommand{\ee}{\end{equation}}
\newcommand{\bed}{\begin{displaymath}}
\newcommand{\eed}{\end{displaymath}}
\newcommand{\bl}{\begin{lem}}
\newcommand{\el}{\end{lem}}
\newcommand{\bp}{\begin{prop}}
\newcommand{\ep}{\end{prop}}
\newcommand{\bt}{\begin{thm}}
\newcommand{\et}{\end{thm}}
\newcommand{\Label}{\label}
\newcommand{\bc}{\begin{cor}}
\newcommand{\ec}{\end{cor}}
\newcommand{\la}{\Label}

\newcommand{\br}{\begin{rem}}
\newcommand{\er}{\end{rem}}
\newcommand{\bd}{\begin{defn}}
\newcommand{\ed}{\end{defn}}

\newenvironment{proof}%
{\begin{sloppypar}\noindent{\bf Proof.}}%
{\hspace*{\fill}$\square$\end{sloppypar}\bigskip}

      \def\dC{{\mathbb C}}

      \def\dR{{\mathbb R}}

      \def\cC{{\mathcal C}}
\def\cD{{\mathcal D}}      
   \def\cH{{\mathcal H}}   
      
   \def\cN{{\mathcal N}}

\def\sp{{\text{\rm sp\,}}}

\begin{document}

\title{Scattering matrices and Weyl functions\thanks{This work was supported by DFG, Grant 1480/2}}

\date{\today}

\author{Jussi Behrndt$^a$,
Mark M. Malamud$^b$, 
Hagen Neidhardt$^c$
}

\maketitle

\begin{quote}
{\small \em
a)Technische Universit\"{a}t Berlin,
Institut f\"{u}r Mathematik,
Stra\ss e des 17.\ Juni 136,
D--10623 Berlin, Germany,\\
E-mail: behrndt@math.tu-berlin.de\\

b) 
Donetsk National University,
Department of Mathematics,
Universitetskaya 24,
83055 Donetsk, Ukraine,\\
E-mail: mmm@telenet.dn.ua\\

c)
Weierstra{\ss}-Institut f\"ur
Angewandte Analysis und Stochastik,
Mohrenstr. 39,
 D-10117 Berlin, Germany,\\
E-mail: neidhardt@wias-berlin.de
}
\vspace{8mm}

\noindent 
{\small Abstract:}

\noindent
For a scattering system $\{A_\Theta,A_0\}$ consisting of
selfadjoint extensions $A_\Theta$ and $A_0$ of a symmetric operator
$A$ with finite deficiency indices, the scattering matrix
$\{S_\gT(\gl)\}$ and a spectral shift function
$\xi_\Theta$ are calculated in terms of the Weyl function associated
with the boundary triplet for $A^*$ and a simple proof of the
Krein-Birman formula is given. The results are applied to singular
Sturm-Liouville operators with scalar and matrix potentials, to
Dirac operators and to Schr\"odinger operators with point
interactions.
\end{quote}

\renewcommand{\thefootnote}{\fnsymbol{footnote}}

\setcounter{footnote}{0}
\renewcommand{\thefootnote}{\arabic{footnote}}

\renewcommand{\thefootnote}{\fnsymbol{footnote}} \setcounter{footnote}{0} %
\renewcommand{\thefootnote}{\arabic{footnote}}

\newpage

\tableofcontents

\section{Introduction}
Let $q\in L^1_{loc}(\dR_+)$ be a real function and consider the
singular Sturm-Liouville differential expression $-\frac{d^2}{dx^2}
+ q$ on $\dR_+$. We assume that $-\frac{d^2}{dx^2} + q$ is in the
limit point case at $\infty$, i.e. the corresponding minimal
operator $L$,
\begin{equation}\label{aop}
Lf=-f^{\prime\prime}+qf,\quad\dom (L)=\bigl\{
f\in\cD_{max}\,:\,f(0)=f^\prime(0)=0\bigr\},
\end{equation}
in $L^2(\dR_+)$ has deficiency indices $(1,1)$. Here $\cD_{max}$
denotes the usual maximal domain consisting of all functions $f\in L^2(\dR_+)$
such that $f$ and $f^\prime$ are locally absolutely continuous and
$-f^{\prime\prime}+qf$ belongs to $L^2(\dR_+)$. It is well-known
that the maximal operator is given by $L^*f=-f^{\prime\prime}+qf$,
$\dom(L^*)=\cD_{max}$, and that all selfadjoint extensions of $L$
in $L^2(\dR_+)$ can be parametrized in the form
\begin{displaymath}
L_\Theta=L^*\upharpoonright\dom(L_\Theta),\,\,\dom(L_\Theta)=\Bigl\{f\in\cD_{max}:
f^\prime(0)=\Theta f(0)\bigr\},\,\Theta\in\overline\dR,
\end{displaymath}
where $\Theta=\infty$ corresponds to the Dirichlet boundary
condition $f(0)=0$.

Since the deficiency indices of $L$ are $(1,1)$ the pair
$\{L_\Theta,L_\infty\}$, $\Theta\in\overline\dR$, performs a
complete scattering system, that is, the wave operators
\begin{equation*}
W_\pm(L_\Theta,L_\infty)=s - \lim_{t\rightarrow\pm\infty}
e^{itL_\Theta}e^{-itL_\infty}P^{ac}(L_\infty)
\end{equation*}
exist and their ranges coincide with the absolutely continuous
subspace $\ran(P^{ac}(L_\Theta))$ of $L_\Theta$, cf.
\cite{BW,Ka1,Wei1,Y}. Here $P^{ac}(L_\infty)$ and $P^{ac}(L_\Theta)$
denote the orthogonal projections onto the absolutely continuous
subspace of $L_\infty$ and $L_\Theta$, respectively. The scattering
operator
$S_\Theta=W_+(L_\Theta,L_\infty)^*W_-(L_\Theta,L_\infty)$
commutes with $L_\infty$ and therefore $S_\Theta$ is
unitarily equivalent to a multiplication operator induced by a
family $\{S_\Theta(\lambda)\}$ of unitary operators
in the spectral representation of $L_\infty$. This family is usually called
the scattering matrix of the scattering system
$\{L_\Theta,L_\infty\}$ and is the most important quantity in the
analysis of scattering processes.

A spectral representation of the selfadjoint realizations of
$-\tfrac{d^2}{dx^2}+q$ and in particular of $L_\infty$ has been
obtained by H.~Weyl in \cite{W1,W2,W3}, see also \cite{LS1,M1}. More
precisely, if $\varphi(\cdot,\lambda)$ and $\psi(\cdot,\lambda)$ are
the fundamental solutions of $-u^{\prime\prime}+qu=\lambda u$
satisfying
\begin{equation*}
\varphi(0,\lambda)=1,\,\,\,\varphi^\prime(0,\lambda)=0\quad\text{and}\quad
\psi(0,\lambda)=0,\,\,\,\psi^\prime(0,\lambda)=1,
\end{equation*}
then there exists a scalar function $m$ such that for each $\lambda\in\dC\backslash\dR$
the function
$x\mapsto \varphi(x,\lambda)+m(\lambda)\psi(x,\lambda)$
belongs to $L^2(\dR_+)$. This so-called Titchmarsh-Weyl
function $m$ is a Nevanlinna function
which admits an integral representation
\begin{equation}\label{1.1}
m(\gl) = \alpha + \int_{-\infty}^\infty \left(\frac{1}{t-\gl} - \frac{t}{1+t^2}\right)d\rho(t)
\end{equation}
with a measure $\rho$ satisfying $\int (1+t^2)^{-1}d\rho(t)<\infty$.
Since $L_\infty$ is unitarily equivalent to the multiplication
operator in $L^2(\dR, d\rho)$ the spectral properties of $L_\infty$
can be completely described with the help of the Borel measure
$\rho$, i.e. $L_\infty$ is absolutely continuous, singular,
continuous or pure point if and only if $\rho$ is so.

It turns out that the scattering matrix $\{S_\Theta(\lambda)\}$ of
the scattering system $\{L_\Theta,L_\infty\}$ and the
Titchmarsh-Weyl function $m$ are connected via
\begin{equation}\label{0.3}
S_\Theta(\lambda)=\frac{\Theta-\overline{m(\lambda+i0)}}{\Theta-m(\lambda+i0)}
\end{equation}
for a.e. $\lambda\in\dR$ with $\imag(m(\lambda+i0))\not=0$, cf. Section~\ref{slops}.
For the special case $q=0$ in \eqref{aop} the Titchmarsh-Weyl function is
given by $m(\lambda)=i\sqrt{\lambda}$,
where $\sqrt{\cdot}$ is defined on $\dC$ with a cut along $\dR_+$ and fixed by $\imag\sqrt{\lambda}>0$
for $\lambda\not\in\dR_+$ and by $\sqrt{\lambda}\geq 0$ for $\lambda\in\dR_+$.
In this case formula \eqref{0.3} reduces to
\begin{equation}\label{1.3}
S_\Theta(\gl) = \frac{\Theta + i\sqrt{\gl}}{\Theta - i\sqrt{\gl}}
\qquad \text{for a.e.}\,\,\gl \in \bR_+
\end{equation}
and was obtained in e.g. \cite[\S 3]{Y}.

The basic aim of the present paper is to generalize the
correspondence \eqref{0.3} between the scattering matrix
$\{S_\Theta(\lambda)\}$ of $\{L_\Theta,L_\infty\}$ and the
Titchmarsh-Weyl function $m$ from above to scattering systems
consisting of a pair of selfadjoint operators, which both are
assumed to be extensions of a symmetric operator with finite
deficiency indices, and an abstract analogon of the function $m$.

For this we use the concept of boundary triplets and associated Weyl
functions developed in \cite{DM91,DM95}. Namely, if $A$ is a densely
defined closed symmetric operator with equal deficiency indices
$n_\pm(A)<\infty$ in a Hilbert space $\gotH$ and
$\Pi=\{\cH,\Gamma_0,\Gamma_1\}$ is a boundary triplet for $A^*$,
then all selfadjoint extensions $A_\Theta$ of $A$ in $\gotH$ are
labeled by the selfadjoint relations $\Theta$ in $\cH$, cf.
Section~\ref{btrips}. The analogon of the Sturm-Liouville operator
$L_\infty$ from above here is the selfadjoint extension
$A_0:=A^*\upharpoonright\ker(\Gamma_0)$ corresponding to
the selfadjoint relation $\{(\begin{smallmatrix} 0\\
h\end{smallmatrix}):h\in\cH\}$. To the boundary triplet $\Pi$ one
associates an operator-valued Nevanlinna function $M$ holomorphic on
$\rho(A_0)$ which admits an integral representation of the form
\eqref{1.1} with an operator-valued measure closely connected with
the spectral measure of $A_0$, see e.g. \cite{ABMN1}. This function $M$
is the abstract analogon of the Titchmarsh-Weyl function $m$ from
above and is called the Weyl function corresponding to the boundary
triplet $\Pi$, cf. Section~\ref{weylreso}.

Since $A$ is assumed to be a symmetric operator with finite
deficiency indices the pair $\{A_\Theta,A_0\}$, where $\Theta$ is an
arbitrary selfadjoint relation in $\cH$, is a complete scattering
system with a corresponding scattering matrix
$\{S_\Theta(\lambda)\}$. Our main result is
Theorem~\ref{scattering}, which states that the direct integral
$L^2(\bR,\mu_L,\kH_\gl)$ performs a spectral representation of the
absolutely continuous part $A^{ac}_0$ of $A_0$ such that the
scattering matrix $\{S_\Theta(\gl)\}$ of the
scattering system $\{A_\Theta,A_0\}$ has the form
\begin{equation}\label{0.5}
S_\Theta(\gl) = I_{\kH_\gl} +
2i\sqrt{\imag(M(\gl))}\bigl(\Theta-M(\gl)\bigr)^{-1}
\sqrt{\imag(M(\gl))}
\end{equation}
for a.e. $\gl \in \bR$, where $M(\gl)
:= M(\gl + i0)$, $\mu_L$ is the Lebesgue measure and $\kH_\gl :=
\ran(\imag(M(\gl)))$. If the Weyl function scalar, i.e. the deficiency
indices of $A$ are $(1,1)$, then we immediately restore \eqref{0.3}
from \eqref{0.5}, see also Corollary~\ref{cor3.8A}. 
We note that in \cite{AP1} (see also \cite{AK1}) V.M.~Adamyan and B.S.~Pavlov have already
obtained a different (unitarily equivalent) expression for the scattering matrix
of a pair of selfadjoint extensions of a symmetric operator with finite
deficiency indices. 

We emphasize that the representation \eqref{0.5} in terms 
of the Weyl function of a fixed boundary triplet has several advantages,
e.g. for Sturm-Liouville operators with matrix potentials, Schr\"{o}dinger
operators with point interactions and Dirac operators the high energy asymptotic of
the scattering matrices can be calculated and explicit formulas 
can be given (see Section~\ref{examples}). 
Furthermore, since the difference of the resolvents of $A_\Theta$ and $A_0$ is a finite
rank operator, the complete scattering system
$\{A_\gT,A_0\}$ admits a so-called spectral shift function
$\xi_\gT$, cf. \cite{K62} and e.g. \cite{BY92a,BY92b}. Recall that $\xi_\Theta$ is a real
function summable with weight $(1 + \gl^2)^{-1}$ such that the trace
formula
\bed \tr\left((A_\gT - z)^{-1} - (A_0 - z)^{-1}\right) = -\int_\bR
\frac{1}{(\gl -z)^2}\,\xi_\gT(\gl)\;d\gl \eed
is valid for $z \in \bC \backslash \bR$. The spectral shift function
is determined by the trace formula up to a real constant. Under the assumption that
$\gT$ is a selfadjoint matrix, we show that the spectral shift function of $\{A_\gT,A_0\}$ is given
(up to a real constant) by
\begin{equation}\label{ssf1}
\xi_\gT(\gl) =
\frac{1}{\pi}\imag\bigl(\tr\left(\log(M(\gl + i 0) - \gT)\right)\bigr)
\quad \text{for a.e.} \quad \gl \in \bR,
\end{equation} 
see Theorem \ref{V.1} and \cite{LSY} for the case $n = 1$. 
With this choice of $\xi_\gT$ 
and the representation \eqref{0.5} of the scattering matrix $\{S_\Theta(\lambda)\}$  
it is easy to prove an analogue of the Birman-Krein formula (see \cite{BK1})
\bed 
\det(S_\gT(\gl)) = \exp\bigl( -2\pi i\xi_\gT(\gl)\bigr)\quad\text{for a.e.}\,\,\lambda\in\dR 
\eed
for scattering systems $\{A_\Theta,A_0\}$ consisting of selfadjoint extensions of a
symmetric operator with finite deficiency indices. Finally we mention that
with the help of the representation \eqref{0.5} in a
forthcoming paper the classical Lax-Phillips scattering theory
will be extended and newly interpreted.

The paper is organized as follows. In Section~\ref{two} we briefly
recall the notion of boundary triplets and associated Weyl functions
and review some standard facts. Section~\ref{scat} is devoted to the
study of scattering systems $\{A_\gT,A_0\}$ consisting of
selfadjoint operators which are extension of a densely defined
closed simple symmetric operator $A$ with finite deficiency indices.
After some preparations we proof the representation \eqref{0.5}
of the scattering matrix $\{S_\Theta(\lambda)\}$ in
Theorem~\ref{scattering}. Section~\ref{ssfunction}
is concerned with the spectral shift function and the Birman-Krein formula.
In Section~\ref{examples} we apply our
general result to singular Sturm-Liouville operators with scalar and
matrix potentials, to Dirac operators and to Schr\"{o}dinger
operators with point interactions. Finally, 
for the convenience of the reader we repeat some basic facts on
direct integrals and spectral representations in the appendix, thus
making our exposition self-contained.

{\bf Notations.}  Throughout the paper $\gotH$ and $\kH$ denote
separable Hilbert spaces with scalar product $(\cdot,\cdot)$. The
linear space of bounded linear operators defined from ${\gotH}$ to
${\kH}$ is denoted by $[{\gotH}, {\kH}]$. For brevity we write
$[\gotH]$ instead of $[\gotH,\gotH]$. The set of closed operators in
$\gotH$ is denoted by $\kC(\gotH)$. By $\widetilde{\kC}(\gotH)$ we
denote the set of closed linear relations in $\gotH$. Notice that
$\kC(\gotH) \subseteq \widetilde{\kC}(\gotH)$. The resolvent set and
the spectrum of a linear operator or relation are denoted by
$\rho(\cdot)$ and $\gs(\cdot)$, respectively. The domain, kernel and
range of a linear operator or relation are denoted by $\dom(\cdot)$,
$\ker(\cdot)$ and $\ran(\cdot)$, respectively. By $\kB(\bR)$ we
denote the Borel sets of $\bR$. The Lebesgue measure on $\kB(\bR)$
is denoted by $\mu_L(\cdot)$.

\section{Extension theory of symmetric operators}\label{two}

\subsection{Boundary triplets and closed extensions}\label{btrips}

Let $A$ be a densely defined closed symmetric operator with equal
deficiency indices $n_\pm(A)=\dim\ker(A^*\mp i)\leq\infty$ in the
separable Hilbert space $\gotH$. We use the concept of boundary
triplets for the description of the closed extensions
$A_\Theta\subset A^*$ of $A$ in $\gotH$, see \cite{DM87,DM91,DM95,GG}.

\begin{defn}
A triplet $\Pi=\{\kH,\gG_0,\gG_1\}$ is called {\rm boundary triplet} for the adjoint
operator $A^*$ if $\kH$ is a Hilbert space and
$\Gamma_0,\Gamma_1:\  \dom(A^*)\rightarrow\kH$ are linear mappings such that

\begin{enumerate}

\item [{\rm (i)}] the abstract second  Green's  identity,
\begin{equation*}
(A^*f,g) - (f,A^*g) = (\gG_1f,\gG_0g) - (\gG_0f,\gG_1g),
\end{equation*}
holds for all $f,g\in\dom(A^*)$ and

\item [{\rm (ii)}] the mapping
$\gG:=(\Gamma_0,\Gamma_1)^\top:  \dom(A^*) \longrightarrow \kH
\times \kH$ is surjective.
\end{enumerate}
\end{defn}

We refer to \cite{DM91} and \cite{DM95} for a detailed study of
boundary triplets and recall only some important facts. First of all
a boundary triplet $\Pi=\{\kH,\gG_0,\gG_1\}$ for $A^*$ exists since
the deficiency indices $n_\pm(A)$ of $A$ are assumed to be equal.
Then $n_\pm(A) = \dim\kH$ holds. We note that a boundary triplet for
$A^*$ is not unique.

An operator $\widetilde A$ is called a {\it proper extension} of $A$
if $\widetilde A$ is closed and satisfies $A\subseteq\widetilde
A\subseteq A^*$. Note that here $A$ is a proper extension of itself. 
In order to describe the set of proper extensions
of $A$ with the help of a boundary triplet
$\Pi=\{\kH,\Gamma_0,\Gamma_1\}$ for $A^*$ we have to consider the
set $\widetilde\kC(\kH)$ of closed linear relations in $\kH$, that
is, the set of closed linear subspaces of $\kH\oplus\kH$. A closed
linear operator in $\kH$ is identified with its graph, so that the
set  $\kC(\kH)$  of closed linear operators in $\kH$ is viewed as a
subset of $\widetilde\kC(\kH).$ For the usual definitions of the
linear operations with linear relations, the inverse, the resolvent
set and the spectrum we refer to \cite{DS87}. Recall that the
adjoint relation $\Theta^*\in\widetilde\kC(\kH)$ of a linear
relation $\Theta$ in $\kH$ is defined as
\begin{equation}\label{thetastar}
\Theta^*:= \left\{ \begin{pmatrix} k\\ k^\prime\end{pmatrix}:
(k,h^\prime)=(k^\prime,h)\,\,\text{for all}\,\,
\begin{pmatrix} h\\ h^\prime\end{pmatrix} \in\Theta\right\}
\end{equation}
and $\Theta$ is said to be {\it symmetric} ({\it selfadjoint}) if
$\Theta\subseteq\Theta^*$ (resp. $\Theta=\Theta^*$). Note that
definition \eqref{thetastar} extends the definition of the adjoint operator.

With a  boundary triplet  $\Pi=\{\kH,\gG_0,\gG_1\}$ for $A^*$ one  associates
two selfadjoint extensions of $A$ defined by
\begin{equation*}
A_0:=A^*\!\upharpoonright\ker(\gG_0)
\quad \text{and}\quad
A_1:=A^*\!\upharpoonright\ker(\gG_1).
\end{equation*}
A description of all proper (closed symmetric, selfadjoint) extensions
of $A$ is given in the next proposition.
Note also that the selfadjointness of $A_0$ and $A_1$ is a
consequence of Proposition \ref{propo} (ii).

\begin{prop}\label{propo}
Let  $\Pi=\{\kH,\gG_0,\gG_1\}$ be a  boundary triplet for  $A^*.$  Then the mapping
\begin{equation}\label{bij}
\Theta\mapsto A_\Theta:= \Gamma^{-1}\Theta=\bigl\{f\in\dom(A^*): \
\Gamma f=(\Gamma_0 f,\Gamma_1 f)^\top \in \Theta\bigr\}
\end{equation}
establishes  a bijective correspondence between the set
$\widetilde\kC(\kH)$ and the set of proper extensions of $A$.
Moreover, for $\Theta\in\widetilde\kC(\kH)$ the following assertions
hold.

\begin{enumerate}

\item [{\rm (i)}] $(A_\Theta)^*=  A_{\Theta^*}$.

\item [{\rm (ii)}] $A_\Theta$ is  symmetric (selfadjoint) if and only if $\Theta$ is
symmetric (resp. selfadjoint).

\item [{\rm (iii)}]
$A_\Theta$ is disjoint with $A_0,$ that is $\dom(A_\Theta)\cap \dom(A_0) =\dom(A),$ if
and only if $\Theta\in \kC(\kH)$. In this case
the extension $A_\Theta$ in \eqref{bij} is given by
\bed
A_\Theta=A^*\!\upharpoonright
\ker\bigl(\Gamma_1-\Theta\Gamma_0\bigr).
\eed
\end{enumerate}
\end{prop}

In the following we shall often be concerned with simple symmetric
operators. Recall that a symmetric operator is said to be {\it
simple} if there is no nontrivial subspace which reduces it to a
selfadjoint operator. By \cite{K49} each symmetric operator $A$ in
$\gotH$ can be written as the direct orthogonal sum $\widehat A\oplus
A_s$ of a simple symmetric operator $\widehat A$ in the Hilbert
space
\begin{equation*}
\widehat\gotH=\clo\spa\bigl\{\ker(A^*-\gl):
\gl\in\bC\backslash\bR\bigr\}
\end{equation*}
and a selfadjoint operator $A_s$ in $\gotH\ominus\widehat\gotH$. Here
$\clospa \{\cdot\}$ denotes the closed linear span of a set.
Obviously $A$ is simple if and only if $\widehat\gotH$ coincides
with $\gotH$.

\subsection{Weyl functions and resolvents of extensions}\label{weylreso}

Let, as in Section \ref{btrips}, $A$ be a densely defined closed
symmetric operator in $\gotH$ with equal deficiency indices. If
$\lambda\in\dC$ is a point of regular type of $A$, i.e.
$(A-\lambda)^{-1}$ is bounded, we denote the {\it defect subspace}
of $A$ by $\kN_\gl=\ker(A^*-\gl)$. The following definition can 
be found in \cite{DM87,DM91,DM95}.

\begin{defn}\label{Weylfunc}
Let $\Pi=\{\kH,\gG_0,\gG_1\}$ be a boundary triplet for $A^*$ and
let $A_0=A^*\!\upharpoonright\ker(\gG_0)$. The operator valued
functions
$\gamma(\cdot) :\ \rho(A_0)\longrightarrow  [\kH,\gotH]$ and  $M(\cdot) :\
\rho(A_0)\longrightarrow  [\kH]$ defined by
\begin{equation}\label{2.3A}
\gamma(\gl):=\bigl(\Gamma_0\!\upharpoonright\kN_\gl\bigr)^{-1} \qquad\text{and}\qquad
M(\gl):=\Gamma_1\gamma(\gl), \quad \gl\in\rho(A_0),
\end{equation}
are called the {\rm $\gamma$-field} and the {\rm Weyl function}, respectively,
corresponding to the boundary triplet $\Pi$.
\end{defn}

It follows from the identity  $\dom(A^*)=\ker(\Gamma_0)\,\dot
+\,\kN_\gl$, $\lambda\in\rho(A_0)$, where as above
$A_0=A^*\!\upharpoonright\ker(\gG_0)$, that the $\gamma$-field
$\gamma(\cdot)$ in \eqref{2.3A} is well defined. It is easily seen
that both  $\gamma(\cdot)$ and $M(\cdot)$ are holomorphic on
$\rho(A_0).$ Moreover, the relations
\begin{equation}\label{gammamu}
\gamma(\mu)=\bigl(I+(\mu-\gl)(A_0-\mu)^{-1}\bigr)\gamma(\gl),
\qquad \gl,\mu\in\rho(A_0),
\end{equation}
and
\begin{equation}\label{mlambda}
M(\gl)-M(\mu)^*=(\gl-\overline\mu)\gamma(\mu)^*\gamma(\gl),
\qquad \gl,\mu\in\rho(A_0),
\end{equation}
are valid (see \cite{DM91}).
The identity \eqref{mlambda} yields that $M(\cdot)$ is a  {\it
Nevanlinna function}, that is, $M(\cdot)$ is holomorphic on $\bC\backslash\bR$ and
takes values in $[\kH]$, $M(\gl)=M(\overline\gl)^*$ for
all $\gl\in\bC\backslash\bR$ and $\imag(M(\gl))$ is a nonnegative operator
for all $\gl$ in the upper half plane $\bC_+=\{\lambda\in\bC:\imag \lambda>0\}$.
Moreover, it follows from \eqref{mlambda}
that $0\in \rho(\imag(M(\gl)))$ holds.
It is important to note that if the operator $A$ is simple, then the
Weyl function $M(\cdot)$ determines the pair $\{A,A_0\}$ uniquely up
to unitary equivalence, cf. \cite{DM87,DM91}.

In the case that the deficiency indices $n_+(A)=n_-(A)$ are finite
the Weyl function $M$ corresponding to $\Pi=\{\cH,\Gamma_0,\Gamma_1\}$
is a matrix-valued Nevanlinna function in the finite dimensional space
$\cH$. From \cite{Don1,Gar1} one gets  the existence of the (strong) limit
\begin{equation*}
M(\lambda+i0)=\lim_{\epsilon\rightarrow +0} M(\lambda+i\epsilon)
\end{equation*}
from the upper half-plane for a.e. $\lambda\in\dR$.

Let now $\Pi=\{\kH,\Gamma_0,\Gamma_1\}$ be a boundary triplet for
$A^*$ with $\gamma$-field $\gamma(\cdot)$ and Weyl function
$M(\cdot)$. The spectrum and the resolvent set of a proper (not
necessarily selfadjoint) extension of $A$ can be described with the
help of the Weyl function. If $A_\Theta\subseteq A^*$ is the
extension corresponding to $\Theta\in\widetilde\kC(\kH)$ via
\eqref{bij}, then a point $\gl\in\rho(A_0)$
($\lambda\in\sigma_i(A_0)$, $i=p,c,r$) belongs to $\rho(A_\Theta)$
if and only if $0\in\rho(\Theta-M(\gl))$ (resp.
$0\in\sigma_i(\Theta-M(\gl))$, $i=p,c,r$). Moreover, for
$\gl\in\rho(A_0)\cap\rho(A_\Theta)$ the well-known resolvent formula
\begin{equation}\label{resoll}
(A_\Theta - \gl)^{-1} = (A_0 - \gl)^{-1} + \gga(\gl)\bigl(\Theta -
M(\gl)\bigr)^{-1}\gga(\overline{\gl})^*
\end{equation}
holds. Formula \eqref{resoll} is a generalization of the known Krein
formula for canonical resolvents. We emphasize that it is valid for
any proper extension of $A$ with a nonempty resolvent set. 
It is worth to note that the Weyl function can also be used to investigate the
absolutely continuous and singular continuous spectrum of extensions
of $A$, cf. \cite{BMN1}.

\section{Scattering matrix and Weyl function}\label{scat}

Throughout this section let $A$ be a densely defined closed
symmetric operator with equal deficiency indices $n_+(A)=n_-(A)$ in
the separable Hilbert space $\gotH$. Let
$\Pi=\{\kH,\Gamma_0,\Gamma_1\}$ be a boundary triplet for $A^*$ and
let $\gamma(\cdot)$ and $M(\cdot)$ be the corresponding
$\gamma$-field and Weyl function, respectively. The selfadjoint
extension $A^*\!\upharpoonright\ker(\Gamma_0)$ of $A$ is denoted by
$A_0$. Let $A_\Theta$ be an arbitrary selfadjoint extension of $A$
in $\gotH$ corresponding to the selfadjoint relation
$\Theta\in\widetilde\kC(\kH)$ via \eqref{bij},
$A_\Theta=A^*\upharpoonright\Gamma^{-1}\Theta$.

Later in this section we will assume
that the deficiency indices of $A$ are finite. In this case the {\it wave operators}
\begin{equation*}
W_\pm(A_\Theta,A_0) := \slim_{t\to\pm\infty}e^{itA_\Theta}e^{-itA_0}P^{ac}(A_0),
\end{equation*}
exist and are complete, where $P^{ac}(A_0)$ denotes the orthogonal
projection onto the absolutely continuous subspace $\gotH^{ac}(A_0)$
of $A_0$. Completeness means that the ranges of
$W_\pm(A_\Theta,A_0)$ coincide with the absolutely continuous
subspace $\gotH^{ac}(A_\Theta)$ of $A_\Theta$, cf. \cite{BW,Ka1,Wei1,Y}. 
The {\it scattering operator} $S_\Theta$ of the {\it scattering system}
$\{A_\Theta,A_0\}$ is then defined by
\begin{equation}\label{st}
S_\Theta:= W_+(A_\Theta,A_0)^*W_-(A_\Theta,A_0).
\end{equation}
Since the scattering operator commutes with $A_0$ it follows 
that it is unitarily equivalent to a multiplication operator
induced by a family $\{S_\Theta(\lambda)\}$ of unitary operators in
a spectral representation of 
$A^{ac}_0:=A_0\upharpoonright \dom(A_0)\cap\gotH^{ac}(A_0)$. The aim of this section is to compute this so-called
{\it scattering matrix} $\{S_\Theta(\lambda)\}$ of the complete scattering system $\{A_\gT,A_0\}$
in a suitable chosen spectral representation of $A^{ac}_0$ in terms
of the Weyl function $M(\cdot)$ and the extension parameter $\gT$,
see Theorem~\ref{scattering}.

For this purpose we introduce the identification operator
\begin{equation}\label{techj}
J:=-(A_\Theta-i)^{-1}(A_0-i)^{-1}\in [\gotH]
\end{equation}
and we set
\begin{equation}\label{bc}
B:=\Gamma_0(A_\Theta+i)^{-1}\quad\text{and}\quad C:=\Gamma_1(A_0-i)^{-1}.
\end{equation}
\begin{lem}
Let $A$ be a densely defined closed symmetric operator in the
separable Hilbert space $\gotH$ and let $\gP = \{\kH,\gG_0,\gG_1\}$
be a boundary triplet for $A^*$. Let $A_0 =
A^*\upharpoonright\ker(\gG_0)$ and let
$A_\Theta=A^*\upharpoonright\Gamma^{-1}\Theta$,
$\Theta\in\widetilde\cC(\cH)$, be a selfadjoint extension of $A$.
Then we have
\begin{equation*}
A_\Theta Jf-JA_0f=(A_\Theta-i)^{-1}f-(A_0-i)^{-1}f, \quad f\in\dom(A_0),
\end{equation*}
and the factorization
\begin{equation}\label{facto}
(A_\Theta-i)^{-1}-(A_0-i)^{-1}=B^*C
\end{equation}
holds, where $B$ and $C$ are given by \eqref{bc}.
\end{lem}
\begin{proof}
The first assertion follows immediately. Let us prove the
factorization \eqref{facto}. If $\gamma(\cdot)$ and $M(\cdot)$
denote the $\gamma$-field and Weyl function, respectively,
corresponding to the boundary triplet $\Pi$, then the resolvent
formula
\begin{equation}\label{reso2}
(A_\Theta - \gl)^{-1} = (A_0 - \gl)^{-1} + \gga(\gl)\bigl(\Theta -
M(\gl)\bigr)^{-1}\gga(\overline{\gl})^*
\end{equation}
holds for all $\lambda\in\rho(A_\Theta)\cap\rho(A_0)$, cf.
\eqref{resoll}. Applying the operator $\Gamma_0$ to \eqref{reso2},
using \eqref{bc}, $A_0=A^*\!\upharpoonright\ker(\Gamma_0)$ and the
relation $\Gamma_0\gamma(-i)=I_{\kH}$ we obtain
\begin{equation*}
\begin{split}
B&=\Gamma_0(A_\Theta+i)^{-1}=\Gamma_0(A_0+i)^{-1}+\Gamma_0\gamma(-i)
\bigl(\Theta-M(-i)\bigr)^{-1}\gamma(i)^*\\
&=\bigl(\Theta-M(-i)\bigr)^{-1}\gamma(i)^*.
\end{split}
\end{equation*}
Hence $\Theta=\Theta^*$ and $M(-i)^*=M(i)$ imply
\begin{equation}\label{bb*}
B^*=\gamma(i)\bigl(\Theta-M(i)\bigr)^{-1}.
\end{equation}
Similarly, setting $A_1 := A^*\!\upharpoonright\ker(\Gamma_1)$ we get
from the resolvent formula \eqref{reso2}
\begin{equation*}
(A_1-i)^{-1}=(A_0-i)^{-1}-\gamma(i) M(i)^{-1}\gamma(-i)^*.
\end{equation*}
On the other hand, by the definition of the Weyl function
$\Gamma_1\gamma(i)=M(i)$ holds. Therefore we obtain
\begin{equation}\label{cc*}
C=\Gamma_1(A_0-i)^{-1}=\gamma(-i)^*\quad\text{and}\quad C^*=\gamma(-i).
\end{equation}
Combining  \eqref{reso2} with  \eqref{bb*} and \eqref{cc*} we
arrive at the factorization \eqref{facto}.
\end{proof}
\begin{lem}\label{blb}
Let $A$ be a densely defined closed symmetric operator in the separable Hilbert
space $\gotH$, let $\gP = \{\kH,\gG_0,\gG_1\}$ be a boundary triplet for $A^*$
and let $M(\cdot)$ be the corresponding Weyl function.
Further, let $A_0 = A^*\upharpoonright\ker(\gG_0)$ and let
$A_\Theta = A^*\!\upharpoonright \gG^{-1}\gT$, $\gT \in \widetilde{\kC}(\kH)$,
be a selfadjoint extension of $A$. Then the
relation
\begin{equation*}
\begin{split}
B(A_\Theta - \gl)^{-1}B^*
=&\frac{1}{1+\gl^2}\bigl(\bigl(\Theta-M(\gl)\bigr)^{-1}
-\bigl(\Theta-M(i)\bigr)^{-1}\bigr) \\
&\qquad -\frac{1}{\gl+i}\,\imag\!\bigl(\Theta-M(i)\bigr)^{-1}
\end{split}
\end{equation*}
holds for all $\gl\in\bC\backslash\{\bR\cup\pm i\}$, where $B$ is given by \eqref{bc}.
\end{lem}
\begin{proof}
By \eqref{bc} we have
\begin{equation*}
B(A_\Theta-\gl)^{-1}B^*=\Gamma_0
\bigl\{\Gamma_0(A_\Theta+i)^{-1}(A_\Theta-\overline\gl)^{-1}(A_\Theta-i)^{-1}\bigr\}^*.
\end{equation*}
It follows from the resolvent formula  \eqref{reso2} that
\begin{equation*}
\Gamma_0(A_\Theta-\mu)^{-1}=\bigl((\Theta-M(\mu)\bigr)^{-1}\gamma(\overline\mu)^*
\end{equation*}
holds for all $\mu\in\bC\backslash\bR.$ Combining this formula with
the identity
\begin{equation*}
\begin{split}
&\quad(A_\Theta + i)^{-1}(A_\Theta - \overline{\gl})^{-1}(A_\Theta - i)^{-1} =\\
&\frac{1}{{\overline{\gl^2}} + 1}
\left\{(A_\Theta -\overline{\gl})^{-1} - (A_\Theta + i)^{-1}\right\} -
\frac{1}{2i(\overline{\gl} - i)}\left\{(A_\Theta - i)^{-1} - (A_\Theta + i)^{-1}\right\}
\end{split}
\end{equation*}
we obtain
\begin{equation*}
\begin{split}
B(A_\Theta-\gl)^{-1}B^*&=\Gamma_0\Biggl\{\frac{1}{\overline{\gl^2}+1}
\Bigl(\bigl(\Theta-M(\overline\gl)\bigr)^{-1}
\gamma(\gl)^*-\bigl(\Theta-M(-i)\bigr)^{-1}\gamma(i)^*\Bigr)\\
-&\frac{1}{2i(\overline{\gl} - i)}\Bigl(\bigl(\Theta-M(i)\bigr)^{-1}\gamma(-i)^*-
\bigl(\Theta-M(-i)\bigr)^{-1}\gamma(i)^*\Bigr)\Biggr\}^*.
\end{split}
\end{equation*}
Calculating the adjoint and making use of
$\Gamma_0\gamma(\mu)=I_{\kH}$, $\mu\in\bC\backslash\bR$, and the symmetry property
$M(\overline\gl) = M(\gl)^*$ the assertion of Lemma \ref{blb}
follows.
\end{proof}
From now on for the rest of this section we will assume that the
deficiency indices $n_+(A)=n_-(A)$ of the symmetric operator $A$ are
finite, $n_{\pm}(A)<\infty$. In this case the dimension of the
Hilbert space $\kH$ in the boundary triplet
$\Pi=\{\kH,\Gamma_0,\Gamma_1\}$ is also finite and coincides with
the number $n_\pm(A)$. Let again
$A_0=A^*\upharpoonright\ker(\Gamma_0)$ and $J$, $B$ and $C$ as in
\eqref{techj} and \eqref{bc}, respectively. Then the operators $BJ$
and $C$ are finite dimensional and hence the linear manifold
\begin{equation}\label{km}
\kM := \spa\bigl\{\ran(P^{ac}(A_0)J^*B^*),\ran(P^{ac}(A_0)
C^*)\bigr\}  \subseteq \gotH^{ac}(A_0)
\end{equation}
is finite dimensional. Therefore there is a spectral core $\Delta_0
\subseteq \sigma_{ac}(A_0)$ of the operator
$A_0^{ac}:=A_0\!\upharpoonright\gotH^{ac}(A_0)$ such that $\kM$ is a
spectral manifold, cf. Appendix~\ref{app}. The spectral measure of
$A_0$ will be denoted by $E_0$. We equip $\kM$ with the semi-scalar
products
\bed
(f,g)_{E_0,\gl}=\frac{d}{d\gl}(E_0(\gl)f,g),\qquad \gl\in\Delta_0,\quad f,g\in\kM,
\eed
and define the finite dimensional Hilbert spaces $\wh{\kM}_\gl$ by
\begin{equation}\label{hatkm}
\wh{\kM}_\gl:=\kM/\ker(\Vert\cdot\Vert_{E_0,\gl}),\qquad
\lambda\in\Delta_0,
\end{equation}
where $\Vert\cdot\Vert_{E_0,\gl}$ is the semi-norm induced by the
semi-scalar product $(\cdot,\cdot)_{E_0,\gl}$, see Appendix
\ref{app}. Further, in accordance with Appendix \ref{app} we
introduce the linear subset $\kD_\gl \subseteq \gotH^{ac}(A_0)$,
$\gl\in\bR$, with the semi-norm  $[\cdot]_{E_0,\gl}$ given by
\eqref{dlambda}. By factorization and completion of $\kD_\gl$ with
respect to the semi-norm $[\cdot]_{E_0,\gl}$ we obtain the Banach
space
\begin{equation*}
\wh{\kD}_\gl
:=\clo_{[\cdot]_{E_0,\gl}}\bigl(\kD_\gl/\ker([\cdot]_{E_0,\gl})\bigr),\quad \gl \in
\bR,
\end{equation*}
where $\clo_{[\cdot]_{E_0,\gl}}$ denotes the completion with
respect to $[\cdot]_{E_0,\gl}$. By $D_\gl:\kD_\gl\rightarrow
\wh{\kD}_\gl$ we denote the canonical embedding operator. From
$\kM\subseteq\kD_\gl$, $\gl\in\Delta_0$, we have
$D_\gl\kM\subseteq\widehat\kD_\gl$. Moreover, since $\kM$ is a
spectral manifold $D_\gl\kM$ coincides with the Hilbert space
$\widehat\kM_\gl$ for every $\gl \in \gD_0$, cf. Appendix \ref{app}.

Following \cite[\S 18.1.4]{BW} we introduce the linear operators $F_{BJ}(\gl)$ and
$F_C(\gl)$ for every $\gl\in\Delta_0$ by
\begin{equation}\label{fbj}
F_{BJ}(\gl) := D_\gl P^{ac}(A_0)J^*B^*\in [\kH,\widehat\kM_\gl]
\end{equation}
and
\begin{equation*}
F_{C}(\gl) := D_\gl P^{ac}(A_0)C^*\in [\kH,\widehat\kM_\gl].
\end{equation*}
\begin{lem}\label{fbjfc}
Let $A$ be a densely defined closed symmetric operator with finite deficiency
indices in the separable Hilbert
space $\gotH$, let $\gP = \{\kH,\gG_0,\gG_1\}$ be a boundary triplet for $A^*$
and let $M(\cdot)$ be the corresponding Weyl function.
Further, let $A_0 = A^*\upharpoonright\ker(\gG_0)$ and let
$A_\Theta = A^*\!\upharpoonright \gG^{-1}\gT$, $\gT \in \widetilde{\kC}(\kH)$,
be a selfadjoint extension of $A$. Then
\begin{equation*}
F_{BJ}(\gl) = -F_C(\gl)
\left\{\frac{1}{\gl + i}\,\imag\!\bigl(\Theta-M(i)\bigr)^{-1}  + \frac{1}{1 + \gl^2}
\bigl(\Theta-M(i)\bigr)^{-1}\right\}
\end{equation*}
and $\widehat\kM_\gl=\ran F_C(\gl)$ holds for all $\gl \in \gD_0$.
\end{lem}
\begin{proof}
Inserting $J$ from \eqref{techj} into \eqref{fbj} we find
\begin{equation*}
F_{BJ}(\gl)=-D_\gl P^{ac}(A_0)(A_0+i)^{-1}(A_\Theta+i)^{-1}B^*.
\end{equation*}
For $f\in\gotH^{ac}(A_0)$ Lemma \ref{varphidl} implies
$D_\gl(A_0+i)^{-1}f=(\gl+i)^{-1}D_\gl f$ and therefore
\begin{equation}\label{fbj2}
\begin{split}
F_{BJ}(\gl)=&-(\gl+i)^{-1}D_\gl P^{ac}(A_0)(A_\Theta+i)^{-1}B^*\\
=&-(\gl+i)^{-1}D_\gl P^{ac}(A_0)\bigl((A_\Theta+i)^{-1}-(A_0+i)^{-1}\bigr)B^*\\
&\qquad -(\gl+i)^{-1}D_\gl P^{ac}(A_0)(A_0+i)^{-1}B^*.
\end{split}
\end{equation}
By  \eqref{mlambda} we have  $2i \gamma(i)^*\gamma(i)=M(i)-M(-i).$
Taking this identity  into account we obtain from  \eqref{reso2},
\eqref{bb*} and \eqref{cc*}
\begin{equation}\label{ll0res}
\begin{split}
\bigl((A_\Theta+i)^{-1}&-(A_0+i)^{-1}\bigr)B^*\\
&\qquad=\gamma(-i)\bigl(\Theta-M(-i)\bigr)^{-1}\gamma(i)^*\gamma(i)
  \bigl(\Theta-M(i)\bigr)^{-1} \\
&\qquad=C^*\bigl(\Theta-M(-i)\bigr)^{-1}\imag
(M(i))\bigl(\Theta-M(i)\bigr)^{-1} \\
&\qquad=C^*\,\imag\!\bigl(\Theta-M(i)\bigr)^{-1}.
\end{split}
\end{equation}
On the other hand, by \eqref{gammamu} we have
$\gamma(i)=(A_0+i)(A_0-i)^{-1}\gamma(-i)$ and this identity
combined with \eqref{cc*} and \eqref{bb*} yields
\begin{equation}\la{b*c*}
B^*=(A_0+i)(A_0-i)^{-1}C^*\bigl(\Theta-M(i)\bigr)^{-1}.
\end{equation}
Inserting \eqref{ll0res} and \eqref{b*c*} into \eqref{fbj2} and making use of
\eqref{cc*}, Lemma \ref{varphidl} and the definition of $F_C(\gl)$ we obtain
\begin{equation*}
\begin{split}
F_{BJ}(\gl)=&-(\gl+i)^{-1}D_\gl P^{ac}(A_0) C^*\,\imag\!\bigl(\Theta-M(i)\bigr)^{-1}\\
&\quad -(\gl^2+1)^{-1}D_\gl P^{ac}(A_0)C^*\bigl(\Theta-M(i)\bigr)^{-1}\\
=&-F_C(\gl)\left\{\frac{1}{\gl + i}\,\imag\!\bigl(\Theta-M(i)\bigr)^{-1}  + \frac{1}{1 +
\gl^2} \bigl(\Theta-M(i)\bigr)^{-1} \right\}
\end{split}
\end{equation*}
for all $\gl\in\Delta_0$. Therefore $\ran F_{BJ}(\gl)\subseteq\ran F_C(\gl)$ and it
follows that $\widehat\kM_\gl$ coincides with $\ran F_C(\gl)$, $\gl\in\Delta_0$. This
completes the proof of Lemma \ref{fbjfc}.
\end{proof}
In the next lemma we show that the spectral manifold $\kM$ defined
by \eqref{km} is generating with respect to $A_0^{ac}$ if the
symmetric operator $A$ is assumed to be simple (cf. Section
\ref{btrips} and \eqref{gener}). The set of all Borel subsets of the
real axis is denoted by $\kB(\bR)$.
\bl\label{VI}
Let $A$ be a densely defined closed symmetric operator in the separable Hilbert
space $\gotH$ and let $A_0$ be a selfadjoint extension of $A$ with spectral measure
$E_0(\cdot)$.
If $A$ is simple, then the condition
\begin{equation}\label{simple3}
\gotH^{ac}(A_0)=\clo\spa\bigl\{E_0(\Delta)f: \Delta\in\kB(\bR),\,f\in\kM\bigr\}
\end{equation}
is satisfied.
\el
\begin{proof}
Since $A$ is assumed to be simple we have $\gotH = \clospa\{\kN_\gl:
\gl \in \bC \backslash \bR\}$, where
$\cN_\lambda=\ker(A^*-\lambda)$. Hence $\gotH^{ac}(A_0) =
\clospa\{P^{ac}(A_0)\kN_\gl: \gl \in \bC \backslash \bR\}$. From
$C^*=\gamma(-i)$ we find $P^{ac}(A_0)\kN_{-i}\subset\kM$ and by
\eqref{gammamu} we have
\begin{equation*}
\kN_\gl = (A_0 + i)(A_0 - \gl)^{-1}\kN_{-i}
\end{equation*}
which yields
\begin{displaymath}
\kN_\gl \subseteq \clospa\bigl\{E_0(\gD)\ran(C^*): \gD \in \kB(\bR)\bigr\}
\end{displaymath}
for $\gl \in \bC \backslash \bR$. Therefore
\begin{displaymath}
P^{ac}(A_0)\kN_\gl \subseteq\clospa\bigl\{E_0(\Delta)P^{ac}(A_0)\ran(C^*):
\Delta\in\kB(\bR)\bigr\} \subseteq \gotH^{ac}(A_0)
\end{displaymath}
for $\gl \in \bC \backslash \bR$. Since $\gotH^{ac}(A_0) =
\clospa\{P^{ac}(A_0)\kN_\gl: \gl \in \bC \backslash \bR\}$ holds we
find \begin{equation*} \gotH^{ac}(A_0) =
\clospa\{E_0(\Delta)P^{ac}(A_0)\ran(C^*): \Delta\in\kB(\bR)\}
\end{equation*}
which proves relation \eqref{simple3}.
\end{proof}
In accordance with Appendix \ref{app} we can perform a direct
integral representation
$L^2(\Delta_0,\mu_L,\widehat\kM_\gl,\kS_\kM)$ of $\gotH^{ac}(A_0)$
with respect to the absolutely continuous part $A^{ac}_0$ of $A_0$,
where $\widehat\kM_\gl$, $\gl\in\Delta_0$, is
defined by \eqref{hatkm}, $\mu_L$ is the Lebesgue measure and
$\kS_\kM$ is the admissible system from Lemma \ref{applem}. We recall
that in this representation $A^{ac}_0$ is
unitarily equivalent to the multiplication operator $M$,
\begin{displaymath}
(M\wh{f})(\gl)  := \gl \wh{f}(\gl), \qquad \wh{f} \in \dom(M),
\end{displaymath}
where
\begin{displaymath}
\dom(M) := \bigl\{\wh{f} \in L^2(\Delta_0,\mu_L,\widehat\kM_\gl,\kS_\kM):\gl\mapsto
\gl\wh{f}(\gl) \in L^2(\Delta_0,\mu_L,\widehat\kM_\gl,\kS_\kM)\bigr\}.
\end{displaymath}
Since the scattering operator $S_\Theta$ (see \eqref{st}) of the
scattering system $\{A_\Theta,A_0\}$ commutes with $A_0$ and
$A_0^{ac}$ Proposition 9.57 of \cite{BW} implies that there exists a
family $\{\widehat S_\Theta (\gl)\}_{\gl\in\Delta_0}$ of unitary
operators in $\{\widehat\kM_\gl\}_{\gl\in\Delta_0}$ such that the
scattering operator $S_\Theta$ is unitarily equivalent to the
multiplication operator $\widehat S_\Theta$ induced by this family
in the Hilbert space $L^2(\Delta_0,\mu_L,\widehat\kM_\gl,\kS_\kM)$.
We note that this family is determined up to a set of Lebesgue
measure zero and is called the {\it scattering matrix}. The
scattering matrix defines the {\it scattering amplitude}
$\{\wh{T}_\Theta(\gl)\}_{\gl\in\Delta_0}$ by
\begin{equation*}
\wh{T}_\Theta(\gl) := \wh{S}_\Theta(\gl) - I_{\widehat\kM_\gl},
\quad \gl \in \gD_0.
\end{equation*}
Obviously, the scattering amplitude induces a multiplication
operator $\wh{T}_\Theta$ in the Hilbert space
$L^2(\Delta_0,\mu_L,\widehat\kM_\gl,\kS_\kM)$ which is unitarily
equivalent to the $T$-{\it operator}
\begin{equation}\label{top}
T_\Theta := S_\Theta - P^{ac}(A_0).
\end{equation}
The scattering amplitude is also determined up to a set of Lebesgue
measure zero. Making use of results from \cite[\S 18]{BW} we
calculate the scattering amplitude of $\{A_\Theta,A_0\}$ in terms of
the Weyl function $M(\cdot)$ and the parameter $\Theta$. Recall that
the limit $M(\lambda+i0)$ exists for a.e. $\lambda\in\dR$, cf.
Section~\ref{weylreso}.
\begin{thm}\label{trep}
Let $A$ be a densely defined closed simple symmetric operator with
finite deficiency indices in the separable Hilbert space $\gotH$,
let $\Pi=\{\kH,\Gamma_0,\Gamma_1\}$ be a boundary triplet for $A^*$
and let $M(\cdot)$ be the corresponding Weyl function. Further, let
$A_0=A^*\!\upharpoonright\ker(\Gamma_0)$ and let
$A_\Theta=A^*\upharpoonright\Gamma^{-1}\Theta$,
$\Theta\in\widetilde\kC(\kH)$, be a selfadjoint extension of $A$.
Then $L^2(\gD_0,\mu_L,\wh{\kM}_\gl,S_\kM)$ is a spectral
representation of $A^{ac}_0$ such that scattering
amplitude $\{\wh{T}_\Theta(\gl)\}_{\gl \in \gD_0}$ of the scattering
system $\{A_\Theta,A_0\}$ admits the representation
\begin{displaymath}
\widehat T_\Theta(\gl)= 2\pi i (1+\gl^2)
F_C(\gl)\bigl(\Theta-M(\gl+i0)\bigr)^{-1}F_C(\gl)^*\in [\widehat\kM_\gl]
\end{displaymath}
for a.e. $\gl \in \gD_0$.
\end{thm}
\begin{proof}
Besides the scattering system $\{A_\Theta,A_0\}$ and the
corresponding scattering operator $S_\Theta$ and $T$-operator
$T_\Theta$ defined in \eqref{st} and \eqref{top}, respectively, we
consider the scattering system $\{A_\Theta,A_0,J\}$, where $J$ is
defined by \eqref{techj}. The wave operators of $\{A_\Theta,A_0,J\}$
are defined by
\begin{equation*}
W_\pm(A_\Theta,A_0;J) := \slim_{t\to\pm\infty}e^{itA_\Theta}Je^{-itA_0}P^{ac}(A_0);
\end{equation*}
they exist and are complete since $A$ has finite deficiency
indices. Note that
\begin{equation}\label{wj}
\begin{split}
W_\pm(A_\Theta,A_0;J)&=-(A_\Theta-i)^{-1}W_\pm(A_\Theta,A_0)(A_0-i)^{-1}\\
&=-W_\pm(A_\Theta,A_0)(A_0 - i)^{-2}
\end{split}
\end{equation}
holds. The scattering operator $S_{J}$ and the $T$-operator $T_J$ of
the scattering system $\{A_\Theta,A_0;J\}$ are defined by
\begin{equation*}
S_J:= W_+(A_\Theta,A_0;J)^*W_-(A_\Theta,A_0;J)
\end{equation*}
and
\begin{equation}\label{tj}
\begin{split}
T_J:&=S_J-W_+(A_\Theta,A_0;J)^*W_+(A_\Theta,A_0;J)\\
&=S_J-(I+A_0^2)^{-2}P^{ac}(A_0),
\end{split}
\end{equation}
respectively. The second equality in \eqref{tj} follows from
\eqref{wj}. Since the scattering operator $S_\Theta$ commutes with
$A_0$ we obtain
\begin{equation}\label{sjs}
S_J = (I + A^2_0)^{-2}S_\Theta
\end{equation}
from \eqref{wj}.
Note that $S_J$ and $T_J$ both commute with $A_0$ and therefore by
\cite[Proposition~9.57]{BW} there are families $\{\widehat S_J(\gl)\}_{\gl\in\Delta_0}$ and $\{\widehat
T_J(\gl)\}_{\gl\in\Delta_0}$ such that the operators $S_J$ and $T_J$ are unitarily
equivalent to the multiplication operators $\widehat S_J$ and $\widehat T_J$ induced by
these families in $L^2(\Delta_0,\mu_L,\widehat\kM_\gl,\kS_\kM)$. From \eqref{st} and
\eqref{tj} we obtain
\begin{equation*}
\widehat T_\Theta(\gl)=\widehat S_\Theta(\gl)-I_{\widehat\kM_\gl}
\quad\text{and}\quad \widehat T_J(\gl)=\widehat S_J(\gl)-\frac{1}
{(1+\gl^2)^{2}}I_{\widehat\kM_\gl}
\end{equation*}
for $\gl\in\Delta_0$. As \eqref{sjs} implies $\widehat
S_J(\gl)=(1+\gl^2)^{-2} \widehat S_\Theta(\gl)$, $\gl\in\Delta_0$,
we conclude
\begin{equation}\label{tjt}
\widehat T_J(\gl)=\frac{1}{(1+\gl^2)^2}\widehat
T_\Theta(\gl),\quad\gl\in\Delta_0.
\end{equation}
In order to apply \cite[Corollary 18.9]{BW} we have to verify
that
\begin{equation}\label{condi}
\lim_{\epsilon\to +0} B(A_\Theta - \gl - i\epsilon)^{-1}B^*
\end{equation}
exists for a.e. $\gl \in \gD_0$ in the operator norm and that
\begin{equation}\label{condii}
\slim_{\gd\to+0}C\left((A_0 - \gl - i\gd)^{-1} - (A_0 - \gl + i\gd)^{-1}\right)f
\end{equation}
exist for a.e. $\gl \in \gD_0$ and all $f\in\kM$, cf.
\cite[Theorem~18.7 and Remark~18.8]{BW}, where $C$ is given by
\eqref{bc}. Since $\kH$ is  a
finite dimensional space it follows from \cite{Don1,Gar1} that the (strong) limit
\begin{equation*}
\lim_{\epsilon\to +0} \Bigl(-\bigl(\Theta-M(\gl+i\epsilon)\bigr)^{-1}\Bigr)=:
-\bigl(\Theta-M(\gl+i0)\bigr)^{-1}
\end{equation*}
of the $[\kH]$-valued Nevanlinna function $\gl\mapsto -(\Theta-M(\gl))^{-1}$ exists
for a.e. $\gl\in\Delta_0$, cf. Section~\ref{weylreso}. Combining this
fact with Lemma \ref{blb} we obtain that
\eqref{condi} holds. Condition
\eqref{condii} is fulfilled since $C$ is a finite dimensional operator and $\kM$ is a
finite dimensional linear manifold. Hence, by \cite[Corollary 18.9]{BW} we have
\begin{equation*}
\widehat T_J(\gl) = 2\pi i\left\{-F_{BJ}(\gl)F_C(\gl)^* +
F_C(\gl)B(A_\Theta - \gl - i0)^{-1}B^*F_C(\gl)^*\right\}
\end{equation*}
for a.e. $\gl \in \gD_0$. Making use of Lemma \ref{fbjfc} and Lemma
\ref{blb} we obtain
\begin{equation*}
\begin{split}
\widehat T_J(\gl)=2\pi i F_C(\gl) \Biggl\{&\frac{1}{\gl +
i}\,\imag\!\bigl(\Theta-M(i)\bigr)^{-1}  + \frac{1}{1 + \gl^2}
\bigl(\Theta-M(i)\bigr)^{-1}\\
&\,\,+ \frac{1}{1+\gl^2}\bigl(\bigl(\Theta-M(\gl+i0)\bigr)^{-1}
-\bigl(\Theta-M(i)\bigr)^{-1}\bigr)\\
&\,\quad -\frac{1}{\gl+i}\,\imag\!\bigl(\Theta-M(i)\bigr)^{-1} \Biggr\}F_C(\gl)^*.
\end{split}
\end{equation*}
Combining this relation with  \eqref{tjt} we conclude
\begin{equation*}
\frac{1}{1+\gl^2}\widehat T_\Theta(\gl)= 2\pi i
F_C(\gl)\bigl(\Theta-M(\gl+i0)\bigr)^{-1} F_C(\gl)^*
\end{equation*}
for a.e. $\gl\in\Delta_0$ which completes the proof.
\end{proof}
In the following we are going to replace the direct integral
$L^2(\gD_0,\mu_L,\wh{\kM}_\gl,\kS_\kM)$ by a more convenient one.
To this end we prove the following lemma.
\bl\label{8}
Let $A$ be a densely defined closed simple symmetric operator with
finite deficiency indices in the separable Hilbert space $\gotH$ and
let $\Pi=\{\kH,\Gamma_0,\Gamma_1\}$ be a boundary triplet for $A^*$
with corresponding Weyl function $M(\cdot)$. Further, let
$A_0=A^*\!\upharpoonright\ker(\Gamma_0)$, let
$A_\Theta=A^*\upharpoonright\Gamma^{-1}\Theta$,
$\Theta\in\widetilde\kC(\kH)$, be a selfadjoint extension of $A$ and
let $\Delta_0$ be a spectral core of $A_0^{ac}$ such that $\kM$ in
\eqref{km} is a spectral manifold. Then
\begin{equation}\label{4.28}
F_C(\gl)^*F_C(\gl) = \frac{1}{\pi(1+\gl^2)}\,\imag (M(\gl+i0))
\end{equation}
holds for a.e. $\gl \in \gD_0$.
\el
\begin{proof}
Let $B$ and $C$ be as in \eqref{bc} and let $\gD_0$ be a spectral
core for $A^{ac}_0$ such that $\kM$ defined by \eqref{km} is a
spectral manifold. By definition of the operator $D_\gl$ we have
\begin{displaymath}
(F_C(\gl)^*F_C(\gl)u,v) =
\frac{d}{d\gl}(E_0(\gl)C^*u,P^{ac}(A_0)C^*v),
\quad u,v \in \kH,
\end{displaymath}
for $\gl \in \gD_0$. It is not difficult to see that
\begin{equation*}
\begin{split}
(E_0(\gd)C^*u,P^{ac}(A_0)C^*v) &=
\int_\gd \frac{d}{d\gl}(E_0(\gl)C^*u,P^{ac}(A_0)C^*v)\,d\mu_L(\lambda)\\
 &=\int_\gd \frac{d}{d\gl}(E_0(\gl)C^*u,C^*v)\, d\mu_L(\lambda)
\end{split}
\end{equation*}
holds for all $u,v\in\kH$ and any Borel set $\gd \subseteq \bR$. Hence, we find
\begin{displaymath}
\frac{d}{d\gl}(E_0(\gl)C^*u,P^{ac}(A_0)C^*v) = \frac{d}{d\gl}(E_0(\gl)C^*u,C^*v)
\end{displaymath}
for a.e. $\gl \in \gD_0$ and $u,v \in \kH$, which yields
\begin{equation*}
\begin{split}
(F_C(&\gl)^*F_C(\gl)u,v)\\
=&\lim_{\gd\to+0} \frac{1}{2\pi i}
\left(\left\{(A_0 - \gl - i\gd)^{-1} - (A_0- \gl + i\gd)^{-1}\right\}C^*u,C^*v\right)
\end{split}
\end{equation*}
for a.e. $\gl \in \gD_0$ and $u,v \in \kH$. From $C=\Gamma_1(A_0-i)^{-1}=\gamma(-i)^*$
(see \eqref{bc} and \eqref{cc*}) and the relation
$\Gamma_1(A_0-\gl)^{-1}=\gamma(\overline\gl)^*$, $\gl\in\bC\backslash\bR$, we obtain
\begin{equation*}
\begin{split}
C\bigl\{(A_0 -& \gl - i\gd)^{-1} - (A_0- \gl + i\gd)^{-1}\bigr\}C^*\\
&=\frac{1}{i-\gl-i\delta}
\bigl\{\gamma(-i)^*\gamma(-i)-\gamma(\gl-i\delta)^*\gamma(-i)\bigr\}\\
&\qquad
-\frac{1}{i-\gl+i\delta}
\bigl\{\gamma(-i)^*\gamma(-i)-\gamma(\gl+i\delta)^*\gamma(-i)\bigr\}.
\end{split}
\end{equation*}
With the help of \eqref{mlambda} it follows that the right hand side
can be written as
\begin{equation*}
\begin{split}
&\frac{1}{i-\gl-i\delta}
\left\{\imag (M(i))+\frac{M(-i)-M(\gl+i\delta)}{i+\gl+i\delta}\right\}\\
&\qquad\quad -\frac{1}{i-\gl+i\delta}\left\{\imag
(M(i))+\frac{M(-i)-M(\gl-i\delta)}{i+\gl-i\delta} \right\}
\end{split}
\end{equation*}
and we conclude
\begin{equation*}
(F_C(\gl)^*F_C(\gl)u,v)=\frac{1}{2\pi i}\frac{1}{1+\gl^2}
\bigl((M(\gl+i0)-M(\gl-i0))u,v\bigr)
\end{equation*}
for a.e. $\gl \in \gD_0$ and $u,v \in \kH$ which immediately yields \eqref{4.28}.
\end{proof}
In order to formulate the main result we introduce the usual Hilbert
spaces $L^2(\Delta_0,\mu_L,\cH)$ and $L^2(\bR,\mu_L,\kH)$ of square
integrable $\kH$-valued functions on the spectral core $\gD_0$ of
$A_0^{ac}$ and on $\dR$, respectively. Note that
$L^2(\gD_0,\mu_L,\kH)$ is subspace of $L^2(\bR,\mu_L,\kH)$. Let us
define the family $\{\kH_\gl\}_{\gl\in\gL^M}$ of Hilbert spaces
$\kH_\gl$ by
\begin{displaymath}
\kH_\gl := \ran\bigl(\imag\left(M(\gl + i0)\right)\bigr) \subseteq \kH,
\quad \gl \in \gL^M,
\end{displaymath}
where $M(\gl + i0) = \lim_{\epsilon\to 0}M(\gl + i\epsilon)$ and
\begin{equation*}
\gL^M := \bigl\{\gl \in \bR: M(\gl + i0) \; \mbox{exists}\bigr\}.
\end{equation*}
We note that $\kH_\gl = \{0\}$ is quite possible and we recall that
$\bR \backslash\gL^M$ has Lebesgue measure zero. By $\{Q(\gl)\}_{\gl
\in \gL^M}$ we denote the family of orthogonal projections from
$\kH$ onto $\kH_\gl$. One easily verifies that the family
$\{Q(\gl)\}_{\gl \in \gL^M}$ is measurable. This family induces an
orthogonal projection $Q_0$,
\bed (Q_0f)(\gl) := Q(\gl)f(\gl), \quad \mbox{for a.e.} \quad \gl
\in \gD_0, \quad f \in L^2(\gD_0,\mu_L,\kH),
\eed
in $L^2(\gD_0,\mu_L,\kH)$. The range of the projection $Q_0$ is
denoted by $L^2(\gD_0,\mu_L,\kH_\gl)$. Similarly, the family
$\{Q(\gl)\}_{\gl \in \gL^M}$ induces an orthogonal projection $Q$ in
$L^2(\bR,\mu_L,\kH)$, the range of $Q$ is denoted by
$L^2(\bR,\mu_L,\kH_\gl)$. We note that $L^2(\gD_0,\mu_L,\kH_\gl)
\subseteq L^2(\bR,\mu_L,\kH_\gl)$ holds.
\bl\label{III.7}
Let $A$ be a densely defined closed simple symmetric operator with
finite deficiency indices in the separable Hilbert space $\gotH$,
let $\Pi= \{\kH,\Gamma_0,\Gamma_1\}$ be a boundary triplet for
$A^*$, $A_0=A^*\!\upharpoonright\ker(\Gamma_0)$, and let $M(\cdot)$
be the corresponding Weyl function. If the Borel set $\gD_0
\subseteq \gs_{ac}(A_0)$ is a spectral core of $A^{ac}_0$, then
$L^2(\gD_0,\mu_L,\kH_\gl) = L^2(\bR,\mu_L,\kH_\gl)$.
\el
\begin{proof}
Define the set $\Lambda_0^M$ by 
\begin{equation}\label{delta0m}
\gL^M_0 := \bigl\{\gl \in \gL^M: \kH_\gl \not= \{0\}\bigr\}.
\end{equation} 
Then we have to verify that 
$\mu_L(\gL^M_0 \backslash \gD_0) = 0$ holds. From
\eqref{mlambda} we obtain
\bed
\imag(M(\gl)) = \imag(\gl)\gga(\gl)^*\gga(\gl), \quad \gl \in \bC_+.
\eed
and from \eqref{gammamu} we conclude that $\imag(M(\gl))$ coincides with
\begin{equation*}
\imag(\lambda)\gga(i)^*
\left\{I + (\overline{\gl} + i)(A_0 - \overline{\gl})^{-1}\right\}
\left\{I + (\gl - i)(A_0 - \gl)^{-1}\right\}
\gga(i).
\end{equation*}
Hence we have
\bed
\imag(M(\gl)) =
\imag(\gl)\gga(i)^*(A_0+i)(A_0 - \overline{\gl})^{-1}
(A_0-i)(A_0 - \gl)^{-1}\gga(i)
\eed
for $\gl \in \bC_+$ and if $\lambda$ tends to $\dR$ from the upper half-plan we get
\bed
\imag(M(\gl)) = \pi (1 + \gl^2)\,\frac{\gga(i)^*E_0(d\gl)\gga(i)}{d\gl}
\eed
for a.e. $\gl \in \bR$. Here $E_0(\cdot)$ is the spectral measure of
$A_0$. Hence for any bounded Borel set $\gd \in
\kB(\bR)$ we obtain
\bed
\int_{\gd} \frac{1}{1+\gl^2}\imag(M(\gl))\,d\mu_L(\lambda) = \pi\gga(i)^*E^{ac}_0(\gd)\gga(i).
\eed
Since $\gD_0$ is a spectral core of $A^{ac}_0$ one
has $E^{ac}_0(\gD_0) = E^{ac}_0(\bR)$ which implies
$E^{ac}_0(\bR\backslash\gD_0) = 0$ and therefore
\bed
\int_{\bR\setminus\gD_0} \frac{1}{1+\gl^2}\imag(M(\gl))\,d\mu_L(\lambda) = 0.
\eed
Hence we have $\imag(M(\gl)) = 0$ for a.e. $\gl \in \bR\backslash\gD_0$ and thus 
$\kH_\gl = \{0\}$ for a.e. $\gl \in \bR\backslash\gD_0$. Consequently 
$\mu_L(\gL^M_0 \backslash \gD_0) = 0$ and Lemma~\ref{III.7} is proved.
\end{proof}
We note that the so-called absolutely continuous closure
${\mbox{cl}}_{ac}(\gL^M_0)$ of the set $\gL^M_0$ (see \eqref{delta0m}),
\bed
{\mbox{cl}}_{ac}(\gL^M_0) := \bigl\{x \in \bR: \mu_L\bigl((x-\epsilon,x+\epsilon)
\cap \gL^M_0\bigr) > 0 \quad \forall \epsilon > 0\bigr\},
\eed
coincides with the absolutely continuous spectrum $\gs_{ac}(A_0)$ of
$A_0$, cf. \cite[Proposition 4.2]{BMN1}.

The following theorem is the main result of this section, we
calculate the scattering matrix of $\{A_\Theta,A_0\}$ in terms of
the Weyl function $M(\cdot)$ and the parameter $\Theta$ in the
direct integral $L^2(\bR,\mu_L,\kH_\gl)$.

\begin{thm}\label{scattering}
Let $A$ be a densely defined closed simple symmetric operator with
finite deficiency indices in the separable Hilbert space $\gotH$ and
let $\Pi= \{\kH,\Gamma_0,\Gamma_1\}$ be a boundary triplet for $A^*$
with corresponding Weyl function $M(\cdot)$. Further, let
$A_0=A^*\!\upharpoonright\ker(\Gamma_0)$  and
let $A_\Theta=A^*\upharpoonright\Gamma^{-1}\Theta$,
$\Theta\in\widetilde\kC(\kH)$, be a selfadjoint extension of $A$.
Then $L^2(\bR,\mu_L,\kH_\gl)$ performs a spectral representation of
$A^{ac}_0$ such that the scattering matrix $\{S_\Theta(\gl)\}_{\gl
\in \bR}$ of the scattering system $\{A_\Theta,A_0\}$ admits the
representation
\begin{equation}\label{scatformula}
S_\Theta(\gl) = I_{\kH_\gl} +
2i\sqrt{\imag(M(\gl))}\bigl(\Theta-M(\gl)\bigr)^{-1}
\sqrt{\imag(M(\gl))}\in [\cH_\lambda]
\end{equation}
for a.e. $\gl \in \bR$, where $M(\gl)
:= M(\gl + i0)$ and $\kH_\gl := \ran(\imag(M(\gl)))$.
\end{thm}
\begin{proof}
From the polar decomposition of $F_C(\gl)\in [\cH,\widehat\kM_\gl]$
we obtain a family of partial isometries $V(\gl)\in
[\widehat\kM_\gl,\kH]$ defined for a.e. $\gl\in\Delta_0$ which map
$\wh{\kM}_\gl =\ran F_C(\gl)$ isometrically onto $\kH_\gl$ such that
\begin{equation*}
V(\gl)F_C(\gl) = \frac{1}{\sqrt{\pi(1 + \gl^2)}}\sqrt{\imag (M(\gl + i0))}
\end{equation*}
holds for a.e. $\gl \in \gD_0$. Let us introduce the admissible system
\begin{displaymath}
\kS := \left\{\sum^n_{l=1}\ga_l(\gl) V(\gl)J_\gl f_l\;\Big| \; f_l \in \kM,
  \; \ga_l \in L^\infty(\gD_0,\mu_L), \; n \in \bN\right\} \subseteq
\mbox{\large X}_{\gl \in \gD_0}\kH_\gl.
\end{displaymath}
Since $V\kS_\kM = \kS$ one easily verifies that the operator
\begin{equation*}
\begin{split}
V:L^2(\gD_0,\mu_L,\wh{\kM}_\gl,&\kS_\kM)\longrightarrow L^2(\gD_0,\mu_L,\kH_\gl,\kS),\\
(V\wh{f})(\gl) &:= V(\gl)\wh{f}(\gl), \quad \gl \in \gD_0,
\end{split}
\end{equation*}
defines an isometry acting from
$L^2(\gD_0,\mu_L,\wh{\kM}_\gl,\kS_\kM)$ onto
$L^2(\gD_0,\mu_L,\kH_\gl,\kS)$ such that the multiplication
operators induced by the independent variable in
$L^2(\gD_0,\mu_L,\wh{\kM}_\gl,\kS_\kM)$ and
$L^2(\gD_0,\mu_L,\kH_\gl,\kS)$ are unitarily equivalent. Hence
$L^2(\gD_0,\mu_L,\kH_\gl,\kS)$ is a spectral representation of
$A^{ac}_0$, too. In the spectral representation
$L^2(\gD_0,\mu_L,\kH_\gl,\kS)$ the operator  $T_\Theta = S_\Theta -
P^{ac}(A_0)$ is unitarily equivalent to the multiplication operator
induced by $\{T_\Theta(\gl)\}_{\gl\in\Delta_0}$,
\begin{displaymath}
T_\Theta(\gl) = V(\gl)\wh{T}_\Theta(\gl)V(\gl)^*, \quad \gl \in
\gD_0,
\end{displaymath}
in $L^2(\gD_0,\mu_L,\kH_\gl,\kS)$. Using Theorem \ref{trep} and
Lemma \ref{8} we find the representation
\begin{displaymath}
T_\Theta(\gl)  =
2i\sqrt{\imag(M(\gl+i0))}\bigl(\Theta-M(\gl+i0)\bigr)^{-1}
\sqrt{\imag(M(\gl+i0))}
\end{displaymath}
for a.e. $\gl \in \gD_0$ and therefore the scattering matrix
$\{S_\Theta(\gl)\}_{\gl \in \gD_0}$ has the form \eqref{scatformula}.

A straightforward computation shows that the direct integral
$L^2(\gD_0,\mu_L,\kH_\gl,\kS)$ is equal to the subspace
$L^2(\gD_0,\mu_L,\kH_\gl) \subseteq L^2(\gD_0,\mu_L,\kH)$. Taking
into account Lemma \ref{III.7} we find $L^2(\gD_0,\mu_L,\kH_\gl,\kS) =
L^2(\bR,\mu_L,\kH_\gl)$
which shows that $L^2(\bR,\mu_L,\kH_\gl)$ performs a spectral
representation of $A^{ac}_0$ such that the scattering matrix is given
by \eqref{scatformula}.
\end{proof}
\begin{rem}\label{detrem}
{\rm 
Note that the scattering matrix $\{S_\Theta(\gl)\}$ in \eqref{scatformula} is defined 
for a.e. $\lambda\in\dR$ and not only on a spectral core of $A_0$. In particular, 
if $\imag(M(\lambda))=0$ for some $\lambda\in\dR$, then 
$\cH_\lambda=\{0\}$ and $S_\Theta(\lambda)=I_{\{0\}}$. In this case we set 
$\det S_\Theta(\lambda)=1$.  
}
\end{rem}

\begin{cor}\label{cor3.8A}
Let $A$, $\Pi$, $A_0$ and $A_\Theta$ be as in Theorem~{\rm
\ref{scattering}} and assume, in addition, that the Weyl function
$M(\cdot)$ is of scalar type, i.e. $M(\cdot)=m(\cdot)I_\cH$ with a
scalar Nevanlinna function $m(\cdot)$. Then $L^2(\bR,\mu_L,\kH_\gl)$
performs a spectral representation of $A^{ac}_0$ such that the
scattering matrix $\{S_\Theta(\gl)\}_{\gl \in \bR}$ of the
scattering system $\{A_\Theta,A_0\}$ admits the representation
\begin{equation*}
S_\Theta(\gl) = I_{\kH_\gl} +
2i\,\imag(m(\gl))\bigl(\Theta-m(\gl)\cdot I_\cH\bigr)^{-1}\in [\cH_\lambda]
\end{equation*}
for a.e. $\gl \in \bR$. Here $\cH_\lambda=\cH$ if $\imag (m(\lambda))\not=0$ and
$\cH_\lambda=\{0\}$ otherwise.
If, in addition $\Theta\in [\kH]$, then
\begin{equation}\label{4.25A}
S_\Theta(\gl)=\bigl(\Theta - \overline{m(\gl)}\cdot
I_\cH\bigr)\bigl(\Theta-m(\gl)\cdot I_\cH\bigr)^{-1}.
\end{equation}
for a.e. $\gl \in \bR$ with $\imag(m(\gl)) \not= 0$.
\end{cor}
\begin{rem}
{\em
It follows from \eqref{scatformula} that if  $\Theta\in [\kH]$, then
the scattering matrix
$\{S_\Theta(\gl)\}$ admits the representation
\begin{equation} \label{4.26B}
 S_\Theta(\gl) =\bigl(\imag (M(\gl))\bigr)^{-1/2}S(\gl)\bigl(\imag (M(\gl))\bigr)^{1/2}\in[\cH_\lambda]
\end{equation}
for a.e. $\lambda\in\dR$ with $\imag(M(\lambda))\not=0$, where
\begin{equation}\label{4.27A}
S(\gl) := \bigl(\Theta-M(\gl-i0)\bigr) \bigl(\Theta-M(\gl+i0)\bigr)^{-1}.
\end{equation}
Here the operator $(\imag (M(\gl)))^{-1/2}$ is well defined in $\cH_\lambda$ for a.e. $\lambda\in\dR$.
It is worth to note that the first (second) factor of $S(\cdot)$ admits a holomorphic
continuation to the lower (resp. upper) half-plane.

If the Weyl function $M(\cdot)=m(\cdot)I_\cH$ is of scalar type and $\Theta\in [\kH]$, then
we have
$S_\Theta(\gl)=S(\gl)$ and relations \eqref{4.26B} and
\eqref{4.27A} turn into \eqref{4.25A}. In this case
$S_\Theta(\cdot)$ itself can be factorized such that both factors
can be continued holomorphically in ${\bC}_-$ and ${\bC}_+$,
respectively.
}
\end{rem}

\section{Spectral shift function}\label{ssfunction}

M.G.~Krein's spectral shift function introduced in \cite{K62} is an important tool in
the spectral and perturbation theory of self-adjoint operators, in particular 
scattering theory. A detailed
review on the spectral shift function can be
found in e.g. \cite{BY92a,BY92b}. Furthermore we mention \cite{GMN1,GM1,GM2}
as some recent papers on the spectral shift function and its various applications. 

Recall that for any pair of selfadjoint operators $H_1,H_0$ in
a separable Hilbert space $\gotH$ such that the resolvents differ by
a trace class operator,
\begin{equation}\label{trace1}
(H_1-\gl)^{-1}-(H_0-\gl)^{-1}\in\gotS_1(\gotH)
\end{equation}
for some (and hence for all) $\lambda\in\rho(H_1)\cap\rho(H_0)$,
there exists a real valued function $\xi(\cdot)\in L^1_{loc}(\bR)$
satisfying the conditions
\begin{equation}\label{shift1}
\tr\left((H_1-\gl)^{-1}-(H_0-\gl)^{-1}\right)= -\int_{\bR}
\frac{1}{(t-\gl)^{2}}\,\xi(t)\,dt,
\end{equation} 
$\gl\in\rho(H_1)\cap\rho(H_0)$, and
\begin{equation}\label{shift2}
\int_{\bR} \frac{1}{1+t^2}\,\xi(t)\, dt <\infty,
\end{equation}
cf. \cite{BY92a,BY92b,K62}.
Such a function $\xi$ is called a {\it spectral shift
function} of the pair $\{H_1,H_0\}$. We emphasize that $\xi$
is not unique, since simultaneously with $\xi$ a function
$\xi+c$, $c\in\bR$, also satisfies both conditions
\eqref{shift1} and \eqref{shift2}. Note that the converse also
holds, namely, any two spectral shift functions for a pair of selfadjoint
operators $\{H_1,H_0\}$ satisfying \eqref{trace1} 
differ by a real constant. We remark that \eqref{shift1} is a
special case of the general formula
\begin{equation}\label{shift1A}
\tr\left(\phi(H_1) - \phi(H_0)\right)= \int_{\bR}
\phi'(t)\,\xi(t)\,dt,
\end{equation}
which is valid for a wide class of smooth functions. A very large
class of such functions $\phi(\cdot)$ has been described in terms of
the Besov classes by V.V.~Peller in \cite{Pe1}.

In Theorem~\ref{V.1} below we find a representation for the spectral shift
function $\xi_\Theta$
of a pair of selfadjoint operators $A_\Theta$ and $A_0$ which are both assumed to be
extensions of a densely defined
closed simple symmetric operator $A$ with finite deficiency
indices. For that purpose we use the definition
\begin{equation}\label{log}
\log (T):=-i\int_0^\infty \bigl((T+it)^{-1}-(1+it)^{-1}I_\cH\bigr)\,dt
\end{equation}
for an operator $T$ on a finite dimensional Hilbert space $\cH$ satisfying
$\imag(T)\geq 0$ and $0\not\in\sigma(T)$, see e.g. \cite{GMN1,Pot}. 
A straightforward calculation shows that
the relation 
\begin{equation}\label{det}
\det(T)=\exp\bigl(\tr\,(\log(T))\bigr)
\end{equation}
holds. Next we choose a special spectral
shift function $\xi_\Theta$ for the pair $\{A_\Theta,A_0\}$ in terms
of the Weyl function $M$ and the parameter $\Theta$, see also
\cite{LSY} for the case of defect one. Making use of Theorem~\ref{scattering} we give a simple
proof of the Birman-Krein formula, cf. \cite{BK1}.
\begin{thm}\label{V.1}
Let $A$ be a densely defined closed simple symmetric operator in the separable Hilbert space
$\gotH$ with finite deficiency indices $n_\pm(A)=n$, let $\Pi=
\{\kH,\Gamma_0,\Gamma_1\}$ be a boundary triplet for $A^*$ and let
$M(\cdot)$ be the corresponding Weyl function. Further, let
$A_0=A^*\!\upharpoonright\ker(\Gamma_0)$  and let
$A_\Theta=A^*\upharpoonright\Gamma^{-1}\Theta$, $\Theta\in [\kH]$,
be a selfadjoint extension of $A$. Then the following holds:
\begin{enumerate}
\item [{\rm (i)}] The limit $\lim_{\epsilon\rightarrow +0}\log(M(\lambda+i\epsilon)-\Theta)$
exists for a.e. $\lambda\in\dR$ and the function
\begin{equation}\label{xitheta}
\xi_\Theta(\lambda):=
\frac{1}{\pi}\imag\bigl(\tr(\log(M(\gl + i0) - \gT))\bigr) 
\quad \text{for a.e.} \,\,\,\gl \in \bR
\end{equation}
is a spectral shift function for the pair $\{A_\Theta,A_0\}$ with $0\leq\xi_\gT(\gl) \le n$.
\item [{\rm (ii)}] The scattering matrix $\{S_\Theta(\lambda)\}_{\lambda\in\bR}$ of the pair 
$\{A_\Theta,A_0\}$
and the spectral shift function $\xi_\Theta$
in \eqref{xitheta} are connected via the Birman-Krein formula 
\begin{equation}\label{5.6a}
\det S_\gT(\lambda)=\exp\bigl(-2\pi i\xi_\Theta(\lambda)\bigr)
\end{equation}
for a.e. $\lambda\in\dR$ (cf. Remark~{\rm \ref{detrem}}).
\end{enumerate}
\end{thm}
\begin{proof}
(i) Since $\lambda\mapsto M(\lambda)-\Theta$ is a Nevanlinna function with values in $[\cH]$ 
and $0\in\rho(\imag(M(\lambda))$ for all $\lambda\in\dC_+$, it
follows that $\log(M(\lambda)-\Theta)$ is well-defined for all $\lambda\in\dC_+$ by \eqref{log}.
According to \cite[Lemma 2.8]{GMN1} the function $\lambda\mapsto \log(M(\lambda)-\Theta)$, 
$\lambda\in\dC_+$, is a $[\cH]$-valued Nevanlinna function such that
\begin{equation*}
0\leq\imag\bigl(\log(M(\lambda)-\Theta)\bigr)\leq \pi I_\cH
\end{equation*}
holds for all $\lambda\in\dC_+$. Hence the limit $\lim_{\epsilon\rightarrow +0}
\log(M(\lambda+i\epsilon)-\Theta)$ exists for a.e. $\lambda\in\dR$ (see \cite{Don1,Gar1} and
Section~\ref{weylreso}) and 
$\lambda\mapsto\tr(\log(M(\lambda)-\Theta))$, $\lambda\in\dC_+$, is a scalar Nevanlinna
function with the property 
\begin{equation*}
0\leq\imag\bigl(\tr(\log(M(\lambda)-\Theta))\bigr)\leq n \pi,\quad\lambda\in\dC_+,
\end{equation*}
that is, the function $\xi_\Theta$ in \eqref{xitheta} satisfies $0\leq\xi_\gT(\gl) \le n$ for 
a.e. $\lambda\in\dR$.

In order to show that \eqref{shift1} holds with $H_1$, $H_0$ and $\xi$ replaced by
$A_\Theta$, $A_0$ and $\xi_\Theta$, respectively, we first verify 
that the relation
\begin{equation}\label{trlog}
\frac{d}{d\lambda}\tr\bigl(\log(M(\lambda)-\Theta)\bigr)=\tr\left((M(\lambda)-\Theta)^{-1}
\frac{d}{d\lambda}M(\lambda)\right) 
\end{equation}
is true for all $\lambda\in\dC_+$. Indeed, for $\lambda\in\dC_+$ we have
\bed
\log(M(\gl) - \gT) = -i\int^\infty_0  \bigl((M(\gl) - \gT + it)^{-1} - (1 + it)^{-1}I_\cH\bigr)\;dt
\eed
by \eqref{log} and this yields 
\bed
\frac{d}{d\gl}\log(M(\gl) - \gT) = 
i\int^\infty_0 (M(\gl) - \gT + it)^{-1}\left(\tfrac{d}{d\lambda}M(\gl)\right)
(M(\gl) - \gT + it)^{-1}dt.
\eed
Hence we obtain
\bed
\frac{d}{d\gl}\,\tr\bigl(\log(M(\gl) - \gT)\bigr) =
i\int^\infty_0 \tr\left((M(\gl) - \gT + it)^{-2}\tfrac{d}{d\lambda}M(\gl)\right)dt
\eed
and since $\frac{d}{dt}(M(\gl) - \gT + it)^{-1} = -i(M(\gl) - \gT + it)^{-2}$ 
for $t\in(0,\infty)$ we conclude
\bed
\frac{d}{d\gl}\tr\bigl(\log(M(\gl) - \gT)\bigr) =
-\int^\infty_0  \frac{d}{dt}\,\tr\left((M(\gl) - \gT + it)^{-1}\tfrac{d}{d\lambda}M(\gl)\right)\; dt
\eed
for all $\gl \in \bC_+$, that is, relation \eqref{trlog} holds.

From \eqref{mlambda} we find
\begin{equation}\label{4.255}
\gamma(\overline\mu)^*\gamma(\gl)=\frac{M(\gl)-M(\overline\mu)^*} {\gl-\mu}, \qquad \gl,
\mu\in\dC\backslash\dR,\,\lambda\not=\mu,
\end{equation}
and passing in \eqref{4.255} to the limit $\mu\to\gl$ one gets
\begin{equation*}\la{5.9}
\gamma(\overline\gl)^*\gamma(\gl)=\frac{d}{d\gl}M(\gl).
\end{equation*}
Making use of formula \eqref{resoll} for canonical resolvents together with \eqref{trlog}
this implies
\begin{equation}\label{5.10}
\begin{split}
\tr\left((A_\Theta-\gl)^{-1}-(A_0-\gl)^{-1}\right)&=
-\tr\left((M(\gl) - \gT)^{-1}\gamma(\overline\gl)^*\gamma(\gl)\right)\\
&=-\frac{d}{d\lambda}\tr\bigl(\log(M(\lambda)-\Theta)\bigr)
\end{split}
\end{equation}
for all $\gl \in \bC_+$. 

Further, by \cite[Theorem 2.10]{GMN1} there exists a $[\cH]$-valued
measurable function $t\mapsto \Xi_\gT(t)$, $t\in\dR$, such that
\bed
\Xi_\gT(t)=\Xi_\gT(t)^*\qquad\text{and}\qquad
0 \le \Xi_\gT(t) \le I_\kH 
\eed
for a.e. $\gl \in \bR$ and the representation
\bed
\log(M(\gl) - \gT) = C + \int_\bR \Xi_\gT(t)\left((t - \gl)^{-1} - t(1 + t^2)^{-1}\right)\,dt,
\quad \gl \in \bC_+,
\eed
holds with some bounded selfadjoint
operator $C$. Hence
\bed
\tr\bigl(\log(M(\gl) - \gT)\bigr) =
\tr(C) + \int_\bR  \tr\left(\Xi_\gT(t)\right)\left((t - \gl)^{-1} - t(1 + t^2)^{-1}\right)\,dt
\eed
for $\gl \in \bC_+$ and we conclude from
\begin{equation*}
\begin{split}
\xi_\Theta(\lambda)&=\lim_{\epsilon\rightarrow +0}
\frac{1}{\pi}\imag\bigl(\tr(\log(M(\gl + i\epsilon) - \gT))\bigr)\\
&=\lim_{\epsilon\rightarrow +0}\frac{1}{\pi}\int_\dR\tr\left(\Xi_\gT(t)\right)
\epsilon\bigl(t-\lambda)^2+\epsilon^2\bigr)^{-1}dt 
\end{split}
\end{equation*}
that $\xi_\gT(\gl) = \tr(\Xi_\gT(\gl))$
is true for a.e. $\gl \in \bR$. Therefore we have 
\begin{equation*}
\frac{d}{d\gl}\tr\bigl(\log(M(\gl) - \gT)\bigr) =
\int_\bR  (t - \gl)^{-2}\xi_\gT(t)\,dt
\end{equation*}
and together with \eqref{5.10} we
immediately get the trace formula 
\begin{equation*}
\tr\left((A_\Theta-\gl)^{-1}-(A_0-\gl)^{-1}\right)= -\int_{\bR}
\frac{1}{(t-\gl)^{2}}\,\xi_\Theta(t)\,dt.
\end{equation*}
The integrability condition \eqref{shift2} holds because of \cite[Theorem~2.10]{GMN1}.
This completes the proof of assertion (i).

(ii) 
To verify the Birman-Krein formula note that by \eqref{det}  
\begin{equation*}
\begin{split}
&\exp\bigl(-2i\imag\bigl(\tr(\log(M(\gl) - \gT))\bigr) \bigr) \\
&\qquad=\exp\bigl(-\tr(\log(M(\gl) - \gT))\bigr)
\exp\bigl(\overline{\tr(\log(M(\gl) - \gT))}\bigr)\\
&\qquad=\frac{\overline{\det(M(\gl) - \gT)}}{\det(M(\gl) - \gT)}
= \frac{\det(M(\gl)^* - \gT)}{\det(M(\gl) - \gT)}
\end{split}
\end{equation*}
holds for all $\lambda\in\dC_+$. Hence we find
\begin{equation}\label{5.13}
\exp\bigl(-2\pi i\xi_\Theta(\lambda)\bigr)= 
\frac{\det\bigl(M(\gl + i0)^* - \gT\bigr)}{\det\bigl(M(\gl + i0) - \gT\bigr)}
\end{equation}
for a.e. $\gl \in \bR$,  where $M(\gl + i0) := \lim_{\epsilon\to+0}M(\gl + i\epsilon)$ 
exists for a.e. $\gl \in \bR$. It follows from the representation
of the scattering matrix in \eqref{scatformula} and the identity $\det(I+A B)=\det(I+B A)$ that
\bea\la{5.14}
\det S(\gl) & = & \det\left(I_{\kH}+2i\bigl(\imag (M(\lambda+i0))\bigr)
                  \bigl(\Theta-M(\gl+i 0)\bigr)^{-1} \right) \nonumber\\
            & = & \det\left(I_{\kH}+\bigl(M(\gl+i 0)-M(\gl+i 0)^*\bigr)
                   \bigl(\Theta-M(\gl+i 0)\bigr)^{-1}\right)   \nonumber\\
            & = & \det\left(\bigl(\Theta-M(\gl+i 0)^*\bigr)\cdot\bigl(\Theta-M(\gl+i 0)\bigr)^{-1}\right) 
                  \nonumber\\
            & = & \frac{\det\bigl(\Theta-M(\gl+i 0)^*\bigr)}{\det\bigl(\Theta-M(\gl+i 0)\bigr)}
\eea
holds for a.e. $\gl \in \bR$. 
Comparing \eqref{5.13} with \eqref{5.14} we obtain \eqref{5.6a}. 
\end{proof}

We note that for singular Sturm-Liouville operators a definition 
for the spectral shift function similar
to \eqref{xitheta} was already used in \cite{GH1}. 

\section{Scattering systems of differential
operators}\label{examples}

In this section the results from Section \ref{scat} and Section~\ref{ssfunction} are illustrated
for some differential operators. In Section \ref{slops} we consider
a Sturm-Liouville differential expression,  in Section~\ref{slopsmat} we
investigate Sturm-Liouville operators with matrix potentials
satisfying certain integrability conditions and
Section~\ref{diracops} deals with scattering systems consisting of
Dirac operators. Finally, Section~\ref{schroe} is devoted to
Schr\"{o}dinger operators with point interactions.
\subsection{Sturm-Liouville operators}\label{slops}

Let $p,q$ and $r$ be real valued functions on $(a,b)$, $-\infty<a<b\leq\infty$, such that
$p(x)\not= 0$ and $r(x)>0$ for a.e.
$x\in (a,b)$ and $p^{-1},q,r\in L^1((a,c))$ for all $c\in (a,b)$.
Moreover we assume that either $b=\infty$ or at least one of the functions $p^{-1},q,r$
does not belong to $L^1((a,b))$. The Hilbert space of all equivalence classes
of measurable functions $f$ defined on $(a,b)$ for which $\vert f\vert^2 r\in L^1((a,b))$
equipped with the usual inner product
\begin{equation*}
(f,g):=\int_a^b f(x)\overline{g(x)} r(x)\, dx
\end{equation*}
will be denoted by $L^2_r((a,b))$. By our assumptions the differential
expression
\begin{equation}\label{diffexp}
\frac{1}{r}\left(-\frac{d}{dx}\left( p\frac{d}{dx}\right)+q\right)
\end{equation}
is regular at the left endpoint $a$ and singular at the right endpoint $b$. In addition we assume
that the limit point case prevails at $b$, that is, the equation
\begin{equation*}
-(pf^\prime)^\prime+qf=\lambda r f,\qquad\lambda\in\dC,
\end{equation*}
has a unique solution $\phi(\cdot,\lambda)$ (up to scalar multiples) in $L^2_r((a,b))$.
We refer to \cite{DS1,Wei1} for sufficient conditions on the coefficients $r,p,q$ such that
\eqref{diffexp} is limit point at $b$.

In $L^2_r((a,b))$ we consider the operator
\begin{equation*}
\begin{split}
(Af)(x):&=\frac{1}{r(x)}\bigl(-(pf^\prime)^\prime(x) +q(x)f(x)\bigr)\\
\dom (A):&=\bigl\{f\in\cD_{max}\,:\, f(a)=(pf^\prime)(a)=0\bigr\},
\end{split}
\end{equation*}
where $\cD_{max}$ denotes the set of all $f\in L^2_r((a,b))$ such
that $f$ and $pf^\prime$ are locally absolutely continuous and
$\tfrac{1}{r}(-(pf^\prime)^\prime+qf)$ belongs to $L^2_r((a,b))$. It
is well known that $A$ is a densely defined closed simple symmetric
operator with deficiency indices $(1,1)$, see e.g. \cite{DS1,Wei1}
and \cite{Gil} for the fact that $A$ is simple. The adjoint operator
$A^*$ is
\begin{equation*}
(A^*f)(x)=\frac{1}{r(x)}\bigl(-(pf^\prime)^\prime(x) +q(x)f(x)\bigr),\qquad
\dom (A^*)=\cD_{max}.
\end{equation*}
If we choose $\Pi=\{\dC,\Gamma_0,\Gamma_1\}$,
\begin{equation*}
\Gamma_0 f:= f(a)\quad\text{and}\quad\Gamma_1 f:=(pf^\prime)(a),\qquad f\in\dom (A^*),
\end{equation*}
then $\gP$ is a boundary triplet for $A^*$ such that the corresponding Weyl function coincides
with the classical Titchmarsh-Weyl coefficient $m(\cdot)$, cf. \cite{Tit1,W1,W2,W3}. In fact, if
$\varphi(\cdot,\lambda)$ and $\psi(\cdot,\lambda)$ denote the fundamental
solutions of the differential equation $-(pf^\prime)^\prime+qf=\lambda r f$ satisfying
\begin{equation*}
\varphi(a,\lambda)=1,\,\,\,(p\varphi^\prime)(a,\lambda)=0\quad\text{and}\quad
\psi(a,\lambda)=0,\,\,\,(p\psi^\prime)(a,\lambda)=1,
\end{equation*}
then $\sp\{\varphi(\cdot,\lambda)+m(\lambda)\psi(\cdot,\lambda)\}=\ker(A^*-\lambda)$,
$\lambda\in\dC\backslash\dR$,
and by applying $\Gamma_0$ and $\Gamma_1$ to the defect elements
it follows that $m(\cdot)$ is the Weyl function corresponding to the
boundary triplet $\Pi$ .

Let us consider the scattering system $\{A_\Theta,A_0\}$, where $A_0:=A^*\upharpoonright \ker(\Gamma_0)$
and
\begin{equation*}
A_\Theta=A^*\upharpoonright\ker(\Gamma_1-\Theta\Gamma_0)=A^*\upharpoonright
\bigl\{f\in\dom (A^*)\,\vert\, (pf^\prime)(a)=\Theta f(a)\bigr\}
\end{equation*}
for some $\Theta\in\dR$.  By Corollary~\ref{cor3.8A} the scattering
matrix has the form
\begin{equation*}
S_\Theta(\lambda)=\frac{\Theta- \overline{m(\gl)}}{\Theta-m(\gl)}
\end{equation*}
for a.e. $\lambda\in\dR$ with $\imag (m(\lambda+i0))\not= 0$, where
$m(\gl) := m(\gl + i0)$, cf. \eqref{0.3}.

Notice, that in the special case $A^*=-d^2/dx^2$, $\dom
(A^*)=W^2_2(\dR_+)$, i.e.
\begin{equation*}
r(x)=p(x)=1,\,\,\, q(x)=0,\,\,\, a=0\,\,\,\text{and}\,\,\,b=\infty,
\end{equation*}
the defect subspaces
$\ker(A^*-\lambda)$, $\lambda\in\dC\backslash\dR$, are spanned by $x\mapsto e^{i\sqrt{\lambda} x}$,
where the square root is defined on $\bC$ with a cut along $[0,\infty)$ and fixed by
$\imag\sqrt{\gl}>0$ for $\lambda\not\in [0,\infty)$ and by $\sqrt{\lambda}\geq 0$ for
$\lambda\in[0,\infty)$. Therefore the Weyl function corresponding to $\Pi$ is
$m(\lambda)=i\sqrt{\lambda}$ and hence the scattering matrix of the scattering system
$\{A_\Theta,A_0\}$ is
\bed
S_\Theta(\lambda)= 1+
2i\sqrt{\lambda}\bigl(\Theta-i\sqrt{\lambda}\bigr)^{-1}
= \frac{\Theta+i\sqrt{\lambda}}{\Theta-i\sqrt{\lambda}},\qquad \lambda\in\dR_+,
\eed
where $\gT \in \bR$, see \cite[\S 3]{Y} and \eqref{1.3}. In this
case the spectral shift function $\xi_\gT(\cdot)$ of the pair $\{A_\Theta,A_0\}$ 
is given by
\begin{equation}\label{sf1}
\xi_\gT(\gl) = 
\begin{cases}
1 - \chi_{[0,\infty)}(\gl)\frac{1}{\pi}\arctan\left(\frac{\sqrt{|\gl|}}{\gT}\right), & \gT > 0,\\
1-\frac{1}{2}\chi_{[0,\infty)}, & \gT=0,\\
\chi_{(-\infty,-\gT^2)}(\gl) - \chi_{[0,\infty)}(\gl)\frac{1}{\pi}\arctan\left(\frac{\sqrt{|\gl|}}{\gT}\right),
& \gT < 0,
\end{cases}
\end{equation}
for a.e. $\gl \in \bR$. 

\subsection{Sturm-Liouville operators with matrix potentials}\label{slopsmat}

Let $Q\in L^\infty(\bR_+,[\dC^n])$ be a matrix valued function such
that $Q(\cdot)=Q(\cdot)^*$ and the functions $x\mapsto Q(x)$ and $x\mapsto xQ(x)$
belong to $L^1(\bR_+,[\dC^n])$. We consider the operator
\begin{equation*}
A:=-\frac{d^2}{dx^2}+ Q,\quad \dom (A):=\bigl\{f\in
W^2_2({\bR}_+,{\bC}^n):\ f(0)=f'(0)=0\bigr\},
\end{equation*}
in $L^2(\bR_+,\dC^n)$. Then $A$ is a densely defined closed simple
symmetric operator with deficiency indices $n_\pm(A)$ both equal to
$n$ and we have $A^*=-d^2/dx^2+Q$, $\dom (A^*)=W^2_2(\bR_+,\bC^n)$.
Setting
\begin{equation}\label{3.29A}
\Gamma_0 f=f(0), \quad    \Gamma_1 f=f'(0), \qquad   f\in   \dom (A^*)=W^2_2(\bR_+,\bC^n),
\end{equation}
we obtain a boundary triplet $\Pi=\{\bC^n,\Gamma_0,\Gamma_1\}$ for $A^*$. Note that
the extension $A_0=A^*\upharpoonright \ker(\Gamma_0)$ corresponds to Dirichlet 
boundary conditions at $0$, 
\begin{equation}\label{a0def}
A_0=-\frac{d^2}{dx^2}+Q,\quad\dom (A_0)=\bigl\{f\in W^2_2(\dR_+,\dC^n)\,:\,
f(0)=0\bigr\}.
\end{equation}

\begin{prop}\la{V.1a}
Let $A=-d^2/dx^2+Q$ and $\Pi$ be as above and denote the corresponding Weyl function by $M(\cdot)$.
Then the following holds.
\begin{enumerate}
\item[{\em (i)}] The function $M(\cdot)$ has poles on
$(-\infty,0)$ with zero as the only possible accumulation point. Moreover,
$M(\cdot)$ admits a continuous continuation from $\dC_+$ onto $\dR_+$ 
and the 
asymptotic relation
\begin{equation}\label{3.41A}
M(\gl+i0) = i\sqrt{\gl}\;I_{\dC^n} + o(1)\quad\text{as}\,\,\gl=\bar\gl \to +\infty
\end{equation}
holds. Here the cut of the square
root $\sqrt{\cdot}$ is along the positive real axis as in Section~{\rm \ref{slops}}.

\item[{\em (ii)}]
If $\gT\in[\dC^n]$ is self-adjoint, then the scattering matrix
$\{S_\gT(\lambda)\}$ of the scattering system $\{A_{\Theta},A_0\}$ behaves
asymptotically like
\begin{equation}\label{3.42A}
S_\Theta(\lambda) = I_{\bC^n} +
2i\sqrt{\gl}\bigl(\Theta-i\sqrt{\gl}\,\cdot I_{\dC^n}\bigr)^{-1} + o(1)
\end{equation}
as $\gl\to +\infty$, which yields $S_\Theta(\gl) \sim -I_{\dC^n}$  as
$\gl\to +\infty$.
\end{enumerate}
\end{prop}
\begin{proof}
(i) Since the spectrum of $A_0$ (see \eqref{a0def}) is discrete in $(-\infty,0)$
with zero as only possible accumulation point (and purely absolutely continuous in
$(0,\infty)$)
it follows that the Weyl function $M(\cdot)$
has only poles in $(-\infty,0)$ possibly accumulating to zero.
To prove the asymptotic properties of $M(\cdot)$ we recall
that under the condition $x\mapsto x Q(x)\in L^1(\bR_+,[\dC^n])$
the equation $A^* y=\gl y$ has an $n\times n$-matrix solution
$E(\cdot,\gl)$ which solves the integral equation
\begin{equation}\label{transoper}
E(x,\gl)= e^{i x\sqrt{\gl}}\;I_{\dC^n} +
\int^{\infty}_x \frac{\sin(\sqrt{\gl}(t-x))}{\sqrt{\gl}}Q(t)E(t,\gl) dt,
\end{equation}
$\gl\in{\overline\bC_+}$, $x \in \bR_+$, see \cite{AgrMar}.
By \cite[Theorem 1.3.1]{AgrMar} the
solution $E(x,\gl)$ is continuous and uniformly bounded for
$\gl\in{\overline\bC_+}$ and $x \in \bR_+$. Moreover, the derivative
$E'(x,\gl)=\tfrac{d}{dx}E(x,\lambda)$ exists, is continuous and
uniformly bounded for $\gl\in{\overline\bC_+}$ and $x \in \bR_+$,
too. From \eqref{transoper} we immediately get the relation
\begin{equation}\label{3.37B}
E(0,\gl)=I_{\dC^n} + \frac{1}{\sqrt{\gl}}o(1) \qquad \text{as} \quad
\real(\gl) \to +\infty, \quad   \gl\in{\overline\bC_+}.
\end{equation}
Since
\bed
E'(x,\gl)= i\sqrt{\gl}e^{i x\sqrt{\gl}}\;I_{\dC^n} -
\int^{\infty}_x \cos(\sqrt{\gl}(t-x))Q(t)E(t,\gl) dt,
\eed
$\gl\in{\overline\bC_+}$, $x \in \bR_+$, we get
\begin{equation}\label{derivasymptotic}
E'(0,\gl)= i\sqrt{\gl}\;I_n + o(1) \qquad \text{as} \quad
\real(\gl) \to+\infty, \quad   \gl\in{\overline\bC_+}.
\end{equation}
In particular, the asymptotic relations \eqref{3.37B} and
\eqref{derivasymptotic} hold as  $\gl\to+\infty$ along the real
axis. Since $A^*E(x,\gl)\xi = \gl E(x,\gl)\xi$, $\xi\in\dC^n$, one gets
\bed
\cN_{\gl}=\ker(A^*-\gl)=\{E(\cdot,\gl)\xi: \   \xi\in\bC^n\},\qquad
\gl\in\bC_+. 
\eed
Therefore using expressions \eqref{3.29A}  for $\Gamma_0$ and
$\Gamma_1$ we obtain
\begin{equation}\label{3.40B}
M(\gl)=E'(0,\gl)\cdot E(0,\gl)^{-1}, \quad \gl\in \bC_+,
\end{equation}
where the existence of $ E(0,\gl)^{-1}$ 
for $\gl\in \bC_+ \cup (0,\infty)$ follows from the surjectivity of the
map $\gG_0$ and the fact that the operator $A_0$ has no eigenvalues in $(0,\infty)$.
Further, by continuity of $E(0,\gl)$ and
$E'(0,\gl)$ in $\gl\in{\overline\bC_+}$ we conclude that the Weyl
function $M(\cdot)$ admits a continuous continuation to $\bR_+$.
Therefore combining  \eqref{3.40B} with  \eqref{3.37B} and
\eqref{derivasymptotic} we arrive at the asymptotic relation
\bed
M(\gl+i0) = E'(0,\gl+i0)\cdot E(0,\gl+i0)^{-1} = i\sqrt{\gl}\;I_{\dC^n} + o(1)
\eed
as $\gl=\bar\gl \to+\infty$ which proves \eqref{3.41A}

(ii) Let now $\Theta=\Theta^*\in [\dC^n]$ and let $A_\Theta=A^*\upharpoonright\ker(\Gamma_1-\Theta\Gamma_0)$ 
be the
corresponding selfadjoint extension of $A$,
\begin{equation*}
A_\Theta=-\frac{d^2}{dx^2}+Q,\quad\dom (A_\Theta)=\left\{f\in W^2_2(\dR_+,\dC^n)\,:\,
\Theta f(0) = f^\prime(0) \right\},
\end{equation*}
and consider the scattering system $\{A_\Theta, A_0\}$, where $A_0$ is given by \eqref{a0def}.
Combining the formula for the scattering matrix $\{S_\Theta(\lambda)\}$,
\begin{equation*}
S_\Theta(\lambda)=I_{\dC^n}+2i\sqrt{\imag(M(\lambda))}\bigl(\Theta-M(\lambda)\bigr)^{-1}
\sqrt{\imag(M(\lambda))}
\end{equation*}
for a.e. $\lambda\in\dR_+$, from Theorem~\ref{scattering} with the asymptotic behaviour
\eqref{3.41A} of the Weyl function $M(\cdot)$ a straightforward calculation 
implies relation \eqref{3.42A} as $\gl\to +\infty$.
Therefore the scattering matrix of the scattering system $\{A_{\Theta},A_0\}$
satisfies $S_\Theta(\gl) \sim -I_{\dC^n}$  as
$\gl\to +\infty$.
\end{proof}
We note that with the help of the asymptotic behaviour \eqref{3.41A} of the Weyl function $M(\cdot)$
also the asymptotic behaviour of the spectral shift function $\xi_\Theta(\cdot)$ of
the pair $\{A_\Theta,A_0\}$ can be calculated. The details are left to the reader.
\begin{rem}
{\em
The high energy asymptotic \eqref{3.42A} is quite
different from the one for the usually considered scattering system
$\{A_0,L_0\}$, where $A_0$ is as in \eqref{a0def},
\bed
L_0 = -\frac{d^2}{dx^2}, \quad \dom(L_0) = \bigl\{f \in
W^2_2(\bR_+,\bC^n): f(0) = 0\bigr\},
\eed
and $Q$ is rapidly decreasing. In this case the scattering matrix
$\{\widetilde S(\gl)\}_{\gl\in\bR_+}$ obeys the relation $\lim_{\gl\to\infty}\widetilde S(\gl)
= I_{\dC^n}$, see \cite{AgrMar}, whereas by Proposition~\ref{V.1a} the scattering matrix 
$\{S_\Theta(\lambda)\}$ of the scattering system $\{A_{\Theta},A_0\}$, $\Theta\in[\dC^n]$ selfadjoint,
satisfies 
$\lim_{\gl\to+\infty}S_\Theta(\lambda) = -I_{\dC^n}$.
}
\end{rem}
Let us now consider the special case $Q=0$. Instead of $A$ and $A^*$ we denote the
minimal and maximal operator by $L$ and $L^*$  and we choose the boundary triplet $\Pi$
from \eqref{3.29A}. Then the defect subspace is
\begin{equation*}
\kN_{\gl}=\bigl\{x\mapsto e^{i\sqrt{\gl}x}\xi:\  \xi\in\bC^n,\, x\in\dR_+\bigr\}, \qquad \gl\in\bC_+\cup
{\bC_-},
\end{equation*}
and the Weyl function $M(\cdot)$ is given by
\begin{equation*}
M(\lambda)=i\sqrt{\lambda}\cdot I_{\dC^n},\qquad \lambda\not\in\dR_+.
\end{equation*}
Let $L_\Theta$ be the selfadjoint extension corresponding to
$\Theta=\Theta^*\in\widetilde\cC(\dC^n)$ and let
$L_0=L^*\upharpoonright\ker\Gamma_0$. By Corollary \ref{cor3.8A} the
scattering matrix $\{S_\Theta(\gl)\}_{\gl\in \bR_+}$ of the
scattering system $\{L_{\Theta},L_0\}$ admits the representation
\begin{equation}\label{3.32}
S_\Theta(\gl)= I_{\dC^n} +2i\sqrt{\lambda}\,
\bigl(\Theta-i\sqrt{\gl}\cdot I_{\dC^n}\bigr)^{-1} \quad\text{for a.e.}\,\,\, \gl\in\bR_+.
\end{equation}
Moreover, if $\Theta\in[\dC^n]$ formula \eqref{3.32} directly yields the asymptotic relation
\begin{equation*}
\lim_{\gl\to+\infty}S_\Theta(\lambda)=-I_{\dC^n}.
\end{equation*} 
If, in particular
$\Theta=0$, then $L_\Theta=L^*\upharpoonright \ker(\Gamma_1)$ is the
operator $-d^2/dx^2$ subject to Neumann boundary conditions
$f'(0)=0$, and we have $S_\Theta(\lambda)=-I_{\dC^n}$,
$\lambda\in\dR_+$.

We note that the spectral shift function $\xi_\gT(\cdot)$ of
the pair $\{L_\gT,L_0\}$ is given by
\begin{equation}\label{sf12}
\xi_\gT(\gl) = \sum^n_{k=1}\xi_{\gT_k}(\gl) \qquad \text{for a.e.}\,\,\gl \in \bR,
\end{equation}
where $\gT_k$, $k =1,2,\ldots,n$, are the eigenvalues of $\Theta=\Theta^*\in[\dC^n]$ and
the functions $\xi_{\gT_k}(\cdot)$ are defined by \eqref{sf1}.

\subsection{Dirac operator}\label{diracops}

Let $a>0$ and let $A$  be a symmetric  Dirac operator on $\bR$ defined
by
\begin{equation*}
\begin{split}
A f&=
\begin{pmatrix} 0&-1\\
1&0
\end{pmatrix}
\frac{d}{dx}f+
\begin{pmatrix} a&0\\
0&-a
\end{pmatrix}
f,\\
\dom (A)&=\left\{f=(f_1, f_2)^\top \in W^1_2(\bR,\bC^2):\ f(0)=0
\right\}.
\end{split}
\end{equation*}
The deficiency indices of $A$ are $(2,2)$ and
$A^*$ is given by
\begin{equation*}
\begin{split}
A^* f&=
\begin{pmatrix} 0&-1\\
1&0
\end{pmatrix}
\frac{d}{dx}f+
\begin{pmatrix} a&0\\
0&-a
\end{pmatrix}
f,\\
\dom (A^*)&=W^1_2(\bR_-,\bC^2)\oplus W^1_2(\bR_+,\bC^2).
\end{split}
\end{equation*}
Moreover, setting
\bed
\Gamma_0 f=\begin{pmatrix}f_2(0-)\\f_1(0+)\end{pmatrix}, \quad
\Gamma_1 f=\begin{pmatrix}f_1(0-)\\f_2(0+)\end{pmatrix}, \quad f=\begin{pmatrix} f_1\\ f_2\end{pmatrix},
\eed
$f_1,f_2\in W^1_2(\bR_-,\bC)\oplus W^1_2(\bR_+,\bC)$,
we obtain a boundary triplet $\Pi=\{\bC^2,\Gamma_0,\Gamma_1\}$ for $A^*$, cf. \cite{BMN1}. Let
the square root $\sqrt{\cdot}$ be defined as in the previous sections and let
$k(\lambda):=\sqrt{\lambda-a}\sqrt{\lambda+a}$, $\lambda\in\dC$. One verifies as in \cite{BMN1}
that $\ker(A^*-\lambda)$, $\lambda\in\dC_+$, is spanned by the
functions
\begin{equation*}
f_{\lambda,\pm}(x):=\begin{pmatrix}\mp i\frac{\sqrt{\lambda+a}}{\sqrt{\lambda-a}}e^{\pm i k(\lambda)x}\\
e^{\pm i k(\lambda)x}
\end{pmatrix}\chi_{\dR_\pm}(x),\quad x\in\dR,\,\lambda\in\dC_+,
\end{equation*}
and hence for $\lambda\in\dC_+$ the Weyl function $M$ corresponding to
the boundary triplet $\Pi$ is given by
\begin{equation}\label{weyldirac}
M(\gl)= \begin{pmatrix}i\sqrt{\tfrac{\gl+a}{\gl-a}} & 0\\ 0 &  i\sqrt{\tfrac{\gl-a}{\gl+a}}\end{pmatrix}
,\qquad
\gl\in\bC_+.
\end{equation}
If $\Theta=\Theta^*$ is a selfadjoint relation in $\dC^2$ and $A_\Theta=A^*\upharpoonright
\Gamma^{-1}\Theta$
is the corresponding extension,
\begin{equation*}
\begin{split}
A_\Theta f&=
\begin{pmatrix} 0&-1\\
1&0
\end{pmatrix}
\frac{d}{dx}f+
\begin{pmatrix} a&0\\
0&-a
\end{pmatrix}
f,\\
\dom (A_\Theta)&=\left\{f=\begin{pmatrix}f_1\\ f_2\end{pmatrix}\in\dom(A^*):
\begin{pmatrix} (f_2(0-),f_1(0+))^\top \\ (f_1(0-),f_2(0+))^\top\end{pmatrix} \in\Theta
\right\},
\end{split}
\end{equation*}
then it follows from Theorem \ref{scattering} that the
scattering matrix $\{S_\Theta(\gl)\}_{\lambda\in\Omega_a}$, where $\Omega_a:=(-\infty,-a)\cup(a,\infty)$, 
of the Dirac scattering system $\{A_\Theta, A_0\}$, $A_0=A^*\upharpoonright\ker(\Gamma_0)$, is given by
\begin{equation}\label{scatformula2}
S_\Theta(\gl) = I_{\dC^2} + 2i\sqrt{\imag(M(\gl))}\bigl(\Theta-M(\gl)\bigr)^{-1}
\sqrt{\imag(M(\gl))}
\end{equation}
for a.e. $\lambda\in\Omega_a$, where
\begin{equation}\label{imdirac}
\imag(M(\gl))= \begin{pmatrix}\sqrt{|\frac{\gl+a}{\gl-a}|} & 0\\
0 & \sqrt{|\frac{\gl-a}{\gl+a}|}\end{pmatrix}, \qquad \gl\in\Omega_a.
\end{equation}
Note that for $\gl\in(-a,a)$ we have $\imag (M(\gl))=0$.

\begin{rem}
{\em
We note that the parameter $\Theta=\Theta^*\in[\dC^2]$, i.e. the boundary
conditions of the perturbed Dirac operator $A_\Theta$, can be
recovered from the limit of the scattering matrix
$S_\Theta(\lambda)$, $\vert\lambda\vert\rightarrow +\infty$,
corresponding to the scattering system $\{A_\Theta,A_0\}$. In fact, it
follows from \eqref{scatformula2}, \eqref{imdirac} and
\eqref{weyldirac} that
\begin{equation*}
S_\Theta(\infty):=\lim_{\vert\lambda\vert\rightarrow +\infty}S_\Theta(\lambda)
=I_{\dC^2}+2i \bigl(\Theta- i\bigr)^{-1}
\end{equation*}
holds. Therefore the
extension parameter $\gT$ is given by 
\begin{equation*}
\Theta = i\bigl(S_\gT(\infty) + I_{\dC^2}\bigr)\bigl(S_\gT(\infty)-I_{\dC^2}\bigr)^{-1}.
\end{equation*}
}
\end{rem}

Assume now that $\Theta=\bigl(\begin{smallmatrix} \theta_1 & 0\\ 0 & \theta_2\end{smallmatrix}\bigr)$,
$\theta_1,\theta_2\in\dR$. Then
\begin{equation*}
\dom (A_\Theta)=\left\{f=\begin{pmatrix}f_1\\
f_2\end{pmatrix}\in\dom(A^*):\,\,
\begin{matrix} \theta_1 f_2(0-)=f_1(0-) \\ \theta_2f_1(0+)=f_2(0+)\end{matrix}
\right\}
\end{equation*}
and the scattering matrix $\{S_\Theta(\lambda)\}_{\lambda\in\Omega_a}$ has the form
\begin{equation*}
S_\Theta(\gl)=\begin{pmatrix} \frac{\theta_1+i\sqrt{|\frac{\gl+a}{\gl-a}|}}
{\theta_1-i\sqrt{\frac{\gl+a}{\gl-a}}} & 0\\ 0 &   \frac{\theta_2+i\sqrt{|\frac{\gl-a}{\gl+a}|}}
{\theta_2-i\sqrt{\frac{\gl-a}{\gl+a}}}
\end{pmatrix},
\quad \gl\in\Omega_a.
\end{equation*}
In this case the spectral shift function $\xi_\Theta$ of the pair $\{A_\Theta,A_0\}$ 
is given by 
\bed
\xi_{\gT}(\gl) = \eta_{\gth_1}(\gl) + \eta_{\gth_2}(\gl) \qquad\text{for a.e.} \,\,
\gl \in \bR,
\eed
where 
\bead
\eta_{\gth_i}(\gl) & := &
\begin{cases}
1 - \chi_{\gO_a}(\gl)\frac{1}{\pi}
\arctan\left(\frac{1}{\gth_i}\sqrt{\left|\frac{\gl + a}{\gl - a}\right|}\right), & \gth_i > 0,\\
1-\frac{1}{2}\chi_{\gO_a}(\gl), & \gth_i=0,\\
\chi_{(\vartheta_i,a)}(\gl) - 
\chi_{\gO_a}(\gl)\frac{1}{\pi}\arctan\left(\frac{1}{\gth_i}\sqrt{\left|\frac{\gl+a}{\gl-a}\right|}\right),
& \gth_i < 0,\end{cases}
\eead 
$i=1,2$, and the real constants $\vartheta_1,\vartheta_2\in (-a,a)$ are given by
\begin{equation*}
\vartheta_1=a \frac{\gth^2_1 - 1}{\gth^2_1 + 1}\qquad\text{and}\qquad 
\vartheta_2= a\frac{1-\gth^2_2}{1+\gth^2_2}.
\end{equation*}

\subsection{Schr\"odinger operators with point
interactions}\label{schroe}

As a further example we consider the matrix Schr\"odinger
differential expression $-\Delta + Q$ in $L^2(\bR^3,\bC^n)$ with a
bounded selfadjoint matrix potential $Q(x)=Q(x)^*$, $x\in \bR^3$.
This expression determines a minimal symmetric operator
\begin{equation}\label{4.40}
H := -\Delta + Q, \quad \dom(H) :=\bigl\{f \in W^2_2(\bR^3,\bC^n):
f(0)=0\bigr\},
\end{equation}
in $L^2(\bR^3,\bC^n)$. Notice that $H$ is closed, since  for any
$x\in \bR^3$ the linear functional $l_{x}:\ f\to f(x)$ is bounded in
$W^2_2(\bR^3,\bC^n)$ due to the Sobolev embedding theorem. Moreover,
it is easily seen that the deficiency indices of $H$ are $n_{\pm}(H)
= n$. We note that if $Q=0$ the self-adjoint extensions of $H$ in 
$L^2(\bR^3,\bC^n)$ are used to model so-called point interactions 
or singular potentials, see e.g. \cite{AGHH88,AK1,BF}.

In the next proposition we define a boundary triplet for the adjoint
$H^*$. For $x=(x_1,x_2,x_3)^\top\in\dR^3$ we agree to write
$r:=\vert x\vert=(x^2_1+x^2_2+x^2_3)^{1/2}$.

\begin{prop}\label{SchrProp}
Let $H$ be the minimal Schr\"odinger operator \eqref{4.40} with a
matrix potential $Q= Q^* \in L^{\infty}(\bR^3,[\bC^n])$. Then the
following assertions hold.
\begin{enumerate}

\item[{\rm  (i)}] The domain of $H^*=-\Delta+Q$ is given by
\begin{equation}\label{4.41}
\dom(H^*) =\left\{f \in L^2(\bR^3,\bC^n):
\ba{c} f= \xi_0\,\frac{e^{-r}}{r}+\xi_1\,e^{-r} + f_H,\\
\xi_0,\xi_1\in\dC^n,\,\, f_H\in \dom(H) \ea \right\}.
\end{equation}

\item[{\rm  (ii)}] A boundary triplet $\Pi =\{\bC^n,\Gamma_0,\Gamma_1\}$ for $H^*$
is defined by
\begin{equation}  \label{4.42}
\Gamma_j f:= 2\sqrt{\pi}\,\xi_j, \quad f=\xi_0\,\frac{e^{-r}}{r} +
\xi_1\,e^{-r} + f_H\in \dom(H^*), \,\,j=0,1.
\end{equation}

\item[{\rm (iii)}] The operator $H_0=H^*\upharpoonright \ker(\Gamma_0)$
is the usual selfadjoint Schr\"odinger operator $-\Delta+Q$ with domain $W^2_2(\dR^3,\dC^n)$.
\end{enumerate}
\end{prop}
\begin{proof}
(i) Since $Q \in L^{\infty}(\bR^3,[\bC^n])$ the domain of $H^*$ does
not depend on $Q$. Therefore it suffices to consider the case $Q=0$.
Here it is well-known, that
\begin{displaymath}
\dom (H^*)=\bigl\{f\in L^2(\dR^3,\dC^n)\cap W^2_{2,\text{\rm
loc}}(\dR^3\backslash\{0\},\dC^n): \Delta f\in
L^2(\dR^3,\dC^n)\bigr\}
\end{displaymath} 
holds, see e.g. \cite{AGHH88,AK1}, and therefore the functions
$x\mapsto e^{-r}/r$ and $x\mapsto e^{-r}$, $r=\vert
x\vert=(x^2_1+x^2_2+x^2_3)^{1/2}$, belong to $\dom(H^*)$. The
linear span of the functions
\begin{displaymath}
x\mapsto \xi_0\frac{e^{-r}}{r}+\xi_1 e^{-r}\qquad \xi_0,\,\xi_1\in\dC^n,
\end{displaymath}
is a $2n$-dimensional subspace in $\dom(H^*)$ and the intersection
with $\dom(H)$ is trivial. Since $\dim(\dom(H^*)/\dom(H)) = 2n$ 
it follows that $\dom(H^*)$ has the form \eqref{4.41}.

(ii) Let $f,g\in\dom(H^*)$. By assertion (i) we have \bed f = h +
f_H, \,\,\, h = \xi_0\,\frac{e^{-r}}{r} + \xi_1\,e^{-r}, \quad
\mbox{and} \quad g = k + g_H, \,\,\, k = \eta_0\,\frac{e^{-r}}{r} +
\eta_1\,e^{-r}, \eed with some functions $f_H, g_H\in\dom(H)$ and
$\xi_0,\xi_1,\eta_0,\eta_1\in\dC^n$. Using polar coordinates we
obtain
\begin{equation*}
\begin{split}
(H^*f,g)& - (f,H^*g) = (H^*h,k) - (h,H^*k)\\
&=4\pi\int^\infty_0 h(r)\frac{\partial}{\partial r}r^2\frac{\partial}{\partial r}\overline{k(r)}\,dr
-4\pi\int^\infty_0 \frac{\partial}{\partial r}r^2\frac{\partial}{\partial r}h(r)\overline{k(r)}\,dr\\
&=4\pi\left[r^2 h(r)\frac{\partial}{\partial r}\overline{k(r)} -
r^2 \frac{\partial}{\partial r}h(r)\overline{k(r)}\right]^{\infty}_{0}
\end{split}
\end{equation*}
and with the help of the relations
\begin{equation*}
r^2\frac{\partial}{\partial r}k(r) = -e^{-r}\left\{(1+r)\eta_0 + r^2\eta_1\right\}
\end{equation*}
and
\begin{equation*}
r^2\frac{\partial}{\partial r}h(r) = -e^{-r}\left\{(1+r)\xi_0 + r^2\xi_1\right\}
\end{equation*}
this implies
\begin{equation*}
\begin{split}
(H^*f,g) - (f,H^*g) = 4\pi \biggl[ e^{-2r}&\left(\xi_0+ r\xi_0 +
r^2\xi_1\right)
\left(\frac{\overline{\eta_0}}{r} + \overline{\eta_1}\right)\\
& - e^{-2r}\left(\frac{\xi_0}{r} + \xi_1\right)
\left(\overline{\eta_0} + r\overline{\eta_0} + r^2\overline{\eta_1}\right)
\biggr]^{\infty}_{0}.
\end{split}
\end{equation*}
This leads to
\bed (H^*f,g) - (f,H^*g) = 4\pi (\xi_1,\eta_0) - 4\pi
(\xi_0,\eta_1)=(\Gamma_1 f,\Gamma_0 g)-(\Gamma_1 f,\Gamma_0g) \eed
and therefore Green's identity  is satisfied.
If follows from \eqref{4.41} that the mapping $\Gamma=(\Gamma_0,\Gamma_1)^\top$ is surjective
and hence assertion ${\rm (ii)}$ is proved.

(iii) Combining \eqref{4.40} and  \eqref{4.41}  we see that any
$f\in W^2_2(\bR^3,\bC^n)$ admits a representation $f= \xi_1
e^{-r}+f_H$ with $\xi_1:= f(0)$ and $f_H = f - \xi_1
e^{-r}\in\dom(H)$ which proves ${\rm (iii)}$.
\end{proof}

It is important to note that the symmetric operator $H$ in
\eqref{4.40} is in general not simple (see e.g. \cite{AGHH88}),
hence $H$ admits a decomposition into a simple part $\widehat{H}$
and a selfadjoint part $H_s$, that is, $H = \widehat{H}\oplus H_s$,
cf. Section~\ref{weylreso}. It is not difficult to see that the
boundary triplet from Proposition \ref{SchrProp} is also a boundary
triplet for $\widehat{H}^*$. Then obviously the Schr\"{o}dinger
operator $H_0$ from Proposition~\ref{SchrProp} (iii) can be written
as $H_0=\widehat H_0\oplus H_s$, where $\widehat H_0=\widehat
H^*\upharpoonright\ker(\Gamma_0)$.

Let us now consider the case where the potential $Q$ is spherically
symmetric, that is, $Q(x)=Q(r)$, $r=(x_1^2+x_2^2+x_3^2)^{1/2}$. In
this case the simple part $\widehat H$ of $H$ becomes unitarily
equivalent to the symmetric Sturm-Liouville operator
\begin{equation*}
A=-\frac{d^2}{dr^2}+Q,\quad \dom(A)=\bigl\{f\in
W^2_2(\dR_+,\dC^n):f(0)=f^\prime(0)=0\bigr\},
\end{equation*}
cf. Section \ref{slopsmat}, and the extension $\widehat H_0$ becomes
unitarily equivalent to the selfadjoint extension $A_0$ of $A$
subject to Dirichlet boundary conditions at $0$.

\begin{prop}\label{ALprop}
Let $H$ be the minimal Schr\"odinger operator with a spherically
symmetric matrix potential $Q=Q^*\in L^\infty(\dR^3,[\dC^n])$ from
\eqref{4.40} and assume that $r\mapsto Q(r)$ and $r\mapsto rQ(r)$
belong to $L^1(\dR_+,[\dC^n])$. Let $\Pi_H$ and $\Pi_A$ be the
boundary triplets for $H^*$ and $A^*$ defined by \eqref{4.42} and
\eqref{3.29A}, respectively. Then the corresponding Weyl functions
$M_{H}(\cdot)$ and $M_A(\cdot)$ are connected via
\begin{equation}
M_H(\lambda)=I_{\dC^n}+M_A(\lambda),\qquad\lambda\in\dC\backslash\dR,
\end{equation}
and the pairs $\{{\widehat H},{\widehat H}_0\}$ and $\{A, A_0\}$ are
unitarily equivalent. If, in particular, $Q=0$,  then $M_H(\gl) =
(i\sqrt{\gl} + 1)\cdot I_{\dC^n}$ .
\end{prop}

\begin{proof}
Let $E(\cdot,\lambda)$, $\lambda\in\dC_+$, be the $n\times n$-matrix
solution of the equation $A^*E(r,\gl)=\gl E(r,\gl)$ from Section
\ref{slopsmat}. Since $E(\cdot,\lambda)\xi\in L^2(\dR_+,[\dC^n])$,
$\xi\in\dC^n$, we see that
\begin{equation*}
U(x,\gl):=\frac{1}{r}E(r,\gl),\qquad r=(x_1^2+x_2^2+x_3^2)^{1/2}\not
=0,
\end{equation*}
satisfies $U(x,\lambda)\xi\in L^2(\dR^3,[\dC^n])$, $\xi\in\dC^n$,
$\lambda\in\dC_+$, and
\begin{equation*}
\begin{split}
H^*U(&x,\lambda)\xi=-\Delta U(x,\lambda)\xi
+Q(r)U(x,\lambda)\xi\\
&=\frac{1}{r}\bigl(-E^{\prime\prime}(\lambda,r)+Q(r)E(r,\lambda)\bigr)\xi=\frac{1}{r}A^*
E(r,\lambda)\xi=\lambda U(x,\lambda)\xi.
\end{split}
\end{equation*}
Therefore $\ker(H^*-\lambda)=\{U(\cdot,\gl)\xi: \xi\in\dC^n\}$,
$\lambda\in\dC_+$. It follows from \eqref{4.42} that
$U(\cdot,\lambda)\xi$ can be decomposed in the form
\begin{equation}\label{4.17}
U(x,\gl)\xi=\frac{1}{r}
E(r,\gl)\xi=\Xi_{0}(\lambda)\xi\,\frac{e^{-r}}{r} + \Xi_1(\gl)\xi
\,e^{-r}+U_H(x,\gl)\xi,
\end{equation} where
  \begin{equation}\label{4.18}
\Xi_0(\lambda)=E(0,\gl),\qquad \Xi_1(\gl)=E(0,\gl) + E'(0,\gl),
   \end{equation}
and $U_H(\cdot,\gl)\in \dom H$.

Note that according to \eqref{3.40B} the Weyl function $M_A(\cdot)$
corresponding to $\Pi_A$ is $M_A(\gl)=E'(0,\gl)\cdot E(0,\gl)^{-1}$,
$\lambda\in\dC_+$ On the other hand, \eqref{4.17} and \eqref{4.18}
imply
\begin{displaymath}
M_H(\gl)=\Xi_1(\gl)\cdot\Xi_0(\gl)^{-1}=\bigl(E(0,\gl)+E'(0,\gl)\bigr)\cdot
E(0,\gl)^{-1}= I_{\dC^n}+M_A(\gl).
\end{displaymath}

The unitary equivalence of the simple operators $\widehat H$ and $A$
as well as of the selfadjoint extensions $\widehat H_0$ and $A_0$ is
a consequence of Corollary 1 and Lemma~2 of \cite{DM91}.
\end{proof}

Let now $H=\widehat H\oplus H_s$ and $Q$ be as in
Proposition~\ref{ALprop} and consider the scattering system
$\{H_\gT,H_0\}$, where
$H_\Theta=H^*\upharpoonright\Gamma^{-1}\Theta$ for some selfadjoint
$\Theta\in\widetilde\cC(\dC^n)$. Then in fact one considers the
scattering system $\{\widehat{H}_\gT,\widehat{H}_0\}$,
$H_\Theta=\widehat H_\Theta\oplus H_s$. In accordance with
Theorem~\ref{scattering} the scattering matrix
$\{\widehat{S}_\gT(\gl)\}_{\gl \in \bR_+}$ of the scattering system
$\{\widehat{H}_\gT,\widehat{H}_0\}$ is given by
\bed \widehat{S}_\gT(\gl) = I_{\dC^n} +
2i\sqrt{\imag(M_A(\lambda))}\bigl(\gT - (M_A(\gl) +
I_{\dC^n})\bigr)^{-1}\sqrt{\imag(M_A(\lambda))} \eed
for a.e. $\gl \in \bR_+$, where $M_A(\cdot)$ is the Weyl function of
the boundary triplet $\Pi_A$, cf. \eqref{3.40B}. If, in particular
$Q=0,$ then $\widehat{S}_\gT(\gl)$ takes the form 
\bed
\widehat{S}_\gT(\gl) = I_{\dC^n} + 2i\sqrt{\gl}\bigl(\gT - (i\sqrt{\gl}
+ 1)\cdot I_{\dC^n}\bigr)^{-1}. 
\eed
In this case the spectral shift function $\widehat{\xi}_\gT(\cdot)$ of the scattering system
$\{\widehat H_\gT,\widehat H_0\}$ is given by
\bed
\widehat{\xi}_\gT(\gl) = \xi_{\gT - I}(\gl) \qquad \text{for a.e.}\,\,\gl \in \bR,
\eed
where $\xi_{\gT - I}(\cdot)$ is the spectral shift function of the scattering system 
$\{L_{\Theta-I},L_0\}$ (see the end of Section~\ref{slopsmat}) defined by \eqref{sf12}.

\begin{appendix}

\section{Direct integrals and spectral representations}\label{app}

Following the lines of \cite{BW} we give a short introduction to
direct integrals of Hilbert spaces and to spectral representations of
selfadjoint operators.

Let $\gL$ be a Borel subset of $\bR$ and let $\mu$ be a Borel measure on
$\bR$. Further, let $\{\kH_\gl,(\cdot,\cdot)_{\cH_\gl}\}_{\gl\in\gL}$ be a family of separable
Hilbert spaces. A subset $\kS$ of the Cartesian product $\mbox{\large X}_{\gl \in
\gL}\kH_\gl$ is called an {\it admissible system} if the following
conditions are satisfied (see \cite{BW}):
\begin{enumerate}

\item The set $\kS$ is linear and $\kS$ is closed with respect to multiplication
by functions from $L^\infty(\gL,\mu)$.

\item For every $f \in \kS$ the function $\gl\mapsto\|f(\gl)\|_{\kH_\gl}$ is Borel
measurable and $\int_\gL \|f(\gl)\|^2_{\kH_\gl}d\mu(\gl) < \infty$.

\item $\spa\{f(\gl)\,\vert\,f \in \kS\}$ is dense in $\kH_\gl$
  (mod $\mu$).

\item If for a Borel subset $\gD \subseteq \gL$ one has
$\int_\gD \|f(\gl)\|^2_{\kH_\gl}d\mu(\gl) = 0$ for all $f \in\kS$,
then $\mu(\gD) = 0$.
\end{enumerate}
A function $f\in\mbox{\large X}_{\gl \in \gL}\kH_\gl$ is {\it strongly
measurable with respect to $\kS$} if there exists a sequence $t_n\in\kS$ such that
$\lim_{n\rightarrow\infty}\Vert f(\gl)-t_n(\gl)\Vert_{\cH_\gl} =0$ (mod $\mu$) is valid. On the
set of all strongly measurable functions $f,g\in\mbox{\large X}_{\gl \in \gL}\kH_\gl$
with the property
\begin{equation*}
\int_\gL\Vert f(\gl)\Vert_{\cH_\gl}^2 d\mu(\gl) < \infty\quad\text{and}\quad \int_\gL\Vert
g(\gl)\Vert_{\cH_\gl}^2 d\mu(\gl) < \infty
\end{equation*}
we introduce the semi-scalar product
\begin{equation*}
(f,g):=\int_\gL\bigl(f(\gl),g(\gl)\bigr)_{\cH_\gl}  d\mu(\gl).
\end{equation*}
By completion of the corresponding factor space one obtains the Hilbert space 
$L^2(\gL,\mu,\kH_\gl,\kS)$ which is
called the {\it direct integral of the family $\kH_\gl$ with respect to $\gL$, $\mu$ and
$\kS$}.

Let in the following $A_0$ be a selfadjoint operator in the separable Hilbert
space $\gotH$, let $E_0$ be the orthogonal spectral measure of
$A_0$, denote the absolutely continuous subspace of $A_0$ by
$\gotH^{ac}(A_0)$ and let $\mu_L$ be the Lebesgue measure.
\begin{defn}
We call a Borel set $\gL \subseteq \sigma_{ac}(A_0)$ a {\rm spectral
core} of the operator $A_0^{ac}:=A_0\upharpoonright \dom(A_0) \cap \gotH^{ac}(A_0)$ if
$E_0(\gL)\gotH^{ac}(A_0)=\gotH^{ac}(A_0)$ and $\mu_L(\Gl)$ is
minimal. A linear manifold $\kM\subseteq \gotH^{ac}(A_0)$ is called
a {\rm spectral manifold} if there exists a spectral core $\gL$ of
$A_0^{ac}$ such that the derivative $\tfrac{d}{d\gl}(E_0(\gl)f,f)$
exists for all $f\in\kM$ and all $\gl\in\gL$.
\end{defn}

Note that every finite dimensional linear manifold $\kM$ in
$\gotH^{ac}(A_0)$ is a spectral manifold. Let us assume that
$\kM\subseteq\gotH^{ac}(A_0)$ is a spectral manifold which is {\it
generating with respect to} $A_0^{ac}$, that is,
\begin{equation}\label{gener}
\gotH^{ac}(A_0)=\clo\spa\bigl\{E_0(\Delta)f: \Delta\in\kB(\bR),\,f\in\kM\bigr\}
\end{equation}
holds and let $\gL$ be a corresponding spectral core of $A_0^{ac}$. We define a family of semi-scalar
products $(\cdot,\cdot)_{E_0,\gl}$ by
\begin{equation*}
(f,g)_{E_0,\gl}:=\frac{d}{d\gl}(E_0(\gl)f,g),\qquad \gl\in\gL,\,f,g\in\kM,
\end{equation*}
and denote the corresponding semi-norms by
$\Vert\cdot\Vert_{E_0,\gl}$. We remark, that the family
$\{(\cdot,\cdot)_{E_0,\gl}\}_{\gl\in\gL}$ is an example of a
so-called {\it spectral form} with respect to the 
spectral measure $E_0^{ac}:=E_0\upharpoonright\gotH^{ac}(A_0)$ of
$A_0^{ac}$ (see \cite[Section~4.5.1]{BW}). By $\widehat\kM_\gl$,
$\gl\in\gL$, we denote the completion of the factor space
\begin{equation*}
\kM\bigr/\ker(\|\cdot\|_{E_0,\gl})
\end{equation*}
with respect to $\|\cdot\|_{E_0,\gl}$. The canonical embedding operator mapping
$\kM$ into the Hilbert space $\widehat\kM_\gl$, $\gl\in\gL$, is
denoted by $J_\gl$,
\begin{equation*}
J_\gl:\kM \rightarrow \widehat\kM_\gl,\quad k\mapsto J_\gl k.
\end{equation*}
\begin{lem}\label{applem}
The set
\begin{displaymath}
\kS_\kM := \Biggl\{\sum^n_{l=1}\ga_l(\gl) J_\gl f_l:  f_l \in \kM, \,\, \ga_l \in
L^\infty(\gL,\mu),\,n\in\bN\Biggr\}\subseteq \mbox{\large X}_{\gl \in \gL}\widehat\kM_\gl
\end{displaymath}
is an admissible system.
\end{lem}
\begin{proof}
Obviously $\kS_\kM$ is linear and closed with respect to multiplication by functions from
$L^\infty(\gL,\mu)$. For $f(\gl)=J_\gl f$, $f\in\kM$, $\gl\in\gL$, we
find from
\begin{equation*}
\Vert f(\gl)\Vert_{\widehat\kM_\gl}^2=\Vert
f\Vert_{E_0,\gl}^2=\frac{d}{d\gl}(E_0(\gl)f,f)
\end{equation*}
that $\gl\mapsto \Vert f(\gl)\Vert_{\widehat\kM_\gl}$ is Borel
measurable and that
\begin{equation*}
\int_\gL\Vert f(\gl)\Vert_{\widehat\kM_\gl}^2 d\mu_L(\gl)=(E_0(\gL)f,f) =(f,f) < \infty
\end{equation*}
holds. Hence it follows that condition (2) is satisfied. For each $\gl\in\gL$ the set
$\{J_\gl f: f\in\kM\}$ is dense in $\widehat\kM_\gl$, thus (3) holds. Finally, if
for some $\Delta\in\kB(\gL)$ and all $f\in\kS_\kM$
\begin{equation*}
0=\int_\Delta \Vert f(\gl)\Vert_{\widehat\kM_\gl}^2 d\mu_L(\gl) =(E_0(\Delta)f,f)=\Vert
E_0(\Delta)f\Vert^2
\end{equation*}
holds, the assumption that $\kM$ is generating implies
$E_0(\Delta)g=0$ for every $g\in\gotH^{ac}(A_0)$, hence
$E_0(\Delta)=0$. As $\gL$ is a spectral core we conclude
$\mu_L(\Delta)=0$.
\end{proof}
Then the direct integral $L^2(\gL,\mu_L,\widehat\kM_\gl,\kS_\kM)$ of
the family $\widehat\kM_\gl$ with respect to the spectral core
$\gL$, the Lebesgue measure and the admissible system $\kS_\kM$ in
Lemma \ref{applem} can be defined. By \cite[Proposition 4.21]{BW}
there exists an isometric operator from $\gotH^{ac}(A_0)$ onto
$L^2(\gL,\mu_L,\widehat\kM_\gl,\kS_\kM)$ such that $E_0(\Delta)$
corresponds to the multiplication operator induced by the
characteristic function $\chi_\Delta$ for any $\Delta\in\kB(\gL)$,
that is, the direct integral
$L^2(\gL,\mu_L,\widehat\kM_\gl,\kS_\kM)$ performs a {\it spectral
representation} of the spectral measure $E_0^{ac}$ of $A_0^{ac}$.

According to \cite[Section 3.5.5]{BW} we introduce the semi-norm $[\cdot]_{E_0,\gl}$,
\begin{equation*}
[f]^2_{E_0,\gl} := \limsup_{h\to 0}\frac{1}{h}\bigl(E_0([\gl,\gl +
h))f,f\bigr), \quad \gl \in \bR, \quad f\ \in \gotH^{ac}(A_0),
\end{equation*}
and we set
\begin{equation}\label{dlambda}
\kD_\gl := \bigl\{f \in \gotH^{ac}(A_0): [f]_{E_0,\gl} <
\infty\bigr\}, \quad \gl \in \bR.
\end{equation}
If $\kM$ is a spectral manifold and $\gL$ is an associated spectral core, then
$\kM \subseteq\kD_\gl$ holds for all $\gl \in \gL$. Moreover, we have
\begin{equation*}
(f,f)_{E_0,\gl} = [f]^2_{E_0,\gl}, \quad f \in \kM, \quad \gl \in \gL.
\end{equation*}
By $\wh{\kD}_\gl$ we denote the Banach space which is obtained from
$\kD_\gl$ by factorization and completion with respect to the
semi-norm $[\cdot]_{E_0,\gl}$, i.e.
\begin{equation*}
\wh{\kD}_\gl := \clo_{[\cdot]_{E_0,\gl}}\bigl(\kD_\gl\bigr/\ker([\cdot]_{E_0,\gl})\bigr).
\end{equation*}
For $\gl \in \gL$ we will regard $\wh{\kM}_\gl$ as a subspace of
$\wh{\kD}_\gl$. By $D_\gl$ we denote the canonical embedding
operator from $\kD_\gl$ into $\wh{\kD}_\gl$. Note that
$\clo\,{D_\gl\kM} = \wh{\kM}_\gl$, $\gl \in \gL$, where the closure
is taken with respect to the topology of $\wh{\kD}_\gl$.
\begin{lem}\label{varphidl}
For a continuous function $\varphi$ on $\sigma(A_0)$ the relation
\begin{equation*}
D_\gl \varphi(A_0)f=\varphi(\gl) D_\gl f
\end{equation*}
holds for all $\gl\in\bR$ and all $f\in\kD_\gl$.
\end{lem}
\begin{proof}
We have to check that
\begin{equation*}
\begin{split}
0&=[\varphi(A_0)f-\varphi(\gl) f]^2_{E_0,\gl}\\
&=\limsup_{h\rightarrow 0}\frac{1}{h}\Bigl(E_0\bigl([\gl,\gl+h)\bigr)
\bigl(\varphi(A_0)-\varphi(\gl)\bigr) f,
\bigl(\varphi(A_0)-\varphi(\gl)\bigr)f\Bigr)\\
&=\limsup_{h\rightarrow 0}\frac{1}{h}\int_{\gl}^{\gl+h}d\Bigl(E_0(t)
\bigl(\varphi(A_0)-\varphi(\gl)\bigr) f, \bigl(\varphi(A_0)-\varphi(\gl)\bigr)f\Bigr)
\end{split}
\end{equation*}
holds for $\gl\in\bR$ and $f\in\kD_\gl$. From
\begin{displaymath}
\Bigl(E_0(t) \bigl(\varphi(A_0)-\varphi(\gl)\bigr) f,
\bigl(\varphi(A_0)-\varphi(\gl)\bigr)f\Bigr)=\int_{-\infty}^t\vert\varphi(s)-\varphi(\gl)\vert^2
d(E_0(s)f,f)
\end{displaymath}
we find
\begin{equation*}
[\varphi(A_0)f-\varphi(\gl) f]^2_{E_0,\gl}=\limsup_{h\rightarrow 0}\frac{1}{h}
\int_{\gl}^{\gl+h}\vert\varphi(t)-\varphi(\gl)\vert^2 d(E_0(t)f,f).
\end{equation*}
As $f$ belongs to $D_\gl$ and $\varphi$ is continuous on
$\sigma(A_0)$ we obtain that this expression is zero.
\end{proof}
\end{appendix}

\end{document}